\documentclass[11pt]{article}
\usepackage[margin=1in]{geometry}
\usepackage{cprotect}
\usepackage{color}
\usepackage[colorlinks=false,linkcolor=blue,citecolor=blue,urlcolor=blue]{hyperref}
\usepackage{amssymb,amsmath}
\usepackage{amsthm}

\usepackage{graphicx}
\usepackage{helvet}
\usepackage[font=footnotesize,labelfont=bf]{caption}
\usepackage{subcaption}
\usepackage{longtable}

\usepackage{multirow}
\usepackage{tabularx}
\usepackage{booktabs}
\usepackage{ragged2e}
\usepackage{diagbox}
\usepackage[demo]{rotating}
\usepackage{authblk}
\usepackage[normalem]{ulem}
\useunder{\uline}{\ul}{}
\usepackage{titlesec}
\usepackage{adjustbox}
\usepackage{textpos}
\usepackage{pifont}
\usepackage{indentfirst}
\usepackage{arydshln}
\usepackage{caption}
\usepackage{afterpage}

\DeclareMathOperator*{\argmin}{arg\,min}
\newcommand*{\tran}{^{\mkern-1.5mu\mathsf{T}}}

\DeclareUnicodeCharacter{2212}{-}

\titleformat*{\section}{\LARGE\bfseries}
\titleformat*{\subsection}{\Large\bfseries}
\titleformat*{\subsubsection}{\large\bfseries}

\usepackage[square,sort&compress,numbers]{natbib}
\makeatletter
\def\blfootnote{\xdef\@thefnmark{}\@footnotetext}
\makeatother

\makeatletter
\def\thinkhline{%
  \noalign{\ifnum0=`}\fi\hrule \@height \thickarrayrulewidth \futurelet
   \reserved@a\@xthickhline}
\def\@xthickhline{\ifx\reserved@a\thickhline
               \vskip\doublerulesep
               \vskip-\thickarrayrulewidth
             \fi
      \ifnum0=`{\fi}}
\makeatother

\newlength{\thickarrayrulewidth}
\usepackage{setspace}
\usepackage{pifont}
\newcommand{\checkcross}{\checkmark\kern-1.1ex\raisebox{.7ex}{\rotatebox[origin=c]{125}{--}}}

\newcommand{\re}{\textcolor{black}}

\numberwithin{equation}{section}
\theoremstyle{plain}

\usepackage{mathtools}

\usepackage[table,dvipsnames]{xcolor}
\usepackage{tabulary}
\usepackage{multirow}
\usepackage{float}
\usepackage{array}
\newcolumntype{L}[1]{>{\raggedright\let\newline\\\arraybackslash\hspace{0pt}}m{#1}}
\newcolumntype{C}[1]{>{\centering\let\newline\\\arraybackslash\hspace{0pt}}m{#1}}
\newcolumntype{R}[1]{>{\raggedleft\let\newline\\\arraybackslash\hspace{0pt}}m{#1}}

\usepackage{amssymb}% http://ctan.org/pkg/amssymb
\usepackage{pifont}% http://ctan.org/pkg/pifont

\newcommand{\R}{{\rm I}\kern-0.18em{\rm R}}
\newcommand{\p}{{\rm I}\kern-0.18em{\rm P}}
\newcommand{\E}{{\rm I}\kern-0.18em{\rm E}}
\newcommand{\1}{{\rm 1}\kern-0.24em{\rm I}}

\newcommand{\beginsupplement}{%
        \setcounter{table}{0}
        \renewcommand{\thetable}{S\arabic{table}}%
        \setcounter{figure}{0}
        \renewcommand{\thefigure}{S\arabic{figure}}%
     }
\newcommand{\quotes}[1]{\textcolor{DarkOrchid}{#1}}

\title{\re{
Categorization and analysis of 14 computational methods for estimating cell potency from single-cell RNA-seq data}}

\author[1]{Qingyang Wang}
\author[1]{Zhiqian Zhai}
\author[2]{Qiuyu Lian}
\author[3]{Dongyuan Song}
\author[1, 3, 4, 5, 6, 7, *]{Jingyi Jessica Li}

\affil[1]{\small Department of Statistics and Data Science, University of California, Los Angeles, CA 90095-1554}
\affil[2]{Wellcome Trust/Cancer Research UK Gurdon Institute, University of Cambridge, Cambridge, Cambridgeshire CB2 1QN}
\affil[3]{Bioinformatics Interdepartmental Ph.D. Program, University of California, Los Angeles, CA 90095-7246}
\affil[4]{Department of Human Genetics, University of California, Los Angeles, CA 90095-7088}
\affil[5]{Department of Computational Medicine, University of California, Los Angeles, CA 90095-1766}
\affil[6]{Department of Biostatistics, University of California, Los Angeles, CA 90095-1772}
\affil[7]{Radcliffe Institute for Advanced Study, Harvard University, Cambridge, MA 02138}
\affil[*]{To whom correspondence should be addressed. Email: jli@stat.ucla.edu}

\date{\vspace{-5ex}}

\begin{document}

\maketitle

\begin{abstract}
In single-cell RNA sequencing (scRNA-seq) data analysis, a critical challenge is to infer hidden cellular dynamic processes from measured static cell snapshots. To tackle this challenge, many computational methods have been developed from distinct perspectives. Besides the common perspectives of inferring pseudotime trajectories and RNA velocities, another important perspective is to estimate the differentiation potential of cells, which is commonly referred to as the ``cell potency." In this review, we provide a comprehensive summary of \re{$14$} computational methods that define cell potency from scRNA-seq data under different assumptions, some of which are even conceptually contradictory. We divide these methods into three categories: \re{average-based}, entropy-based, and correlation-based methods, depending on how a method summarizes gene expression levels of a cell or cell type into a potency measure. Our review focuses on the key similarities and differences of the methods within each category and between the categories, providing a high-level intuition for each method. Moreover, we use a unified set of mathematical notations to detail the \re{$14$} methods' methodologies and summarize their usage complexities, including the number of parameters, the number of required inputs, and the \re{changes from} the method description in publications to the method implementation in software packages. \re{In conclusion, we regard cell potency estimation as an open question that lacks a consensus on the optimal approach, calling for benchmark datasets and studies. Through categorization and detailed description of existing methods, we aim to provide a solid foundation for future benchmark studies of cell potency estimation methods. A broader open challenge is to understand the comparative advantages of methods that infer cellular dynamics from scRNA-seq data using diverse perspectives, including pseudotime trajectories, RNA velocities, and cell potency.} %Realizing the conceptual contradictions of existing methods and the difficulty for fair benchmarking without single-cell-level ground truths, we conclude that accurate estimation of cell potency from scRNA-seq data remains an open challenge. %and also outline the complexities of the methods %In short, most methods define cell potency based on the entropy of a certain distribution of gene expression, and several other methods define cell potency based on some prior knowledge of genes such as ... .

\end{abstract}
\clearpage

\section*{Introduction}

The development of single-cell RNA sequencing (scRNA-seq) enabled researchers to profile gene expression at an unprecedentedly single-cell resolution. However, scRNA-seq provides solely a static snapshot of individual cells, rather than the underlying dynamic process of cell differentiation \re{(i.e., the process during which young, immature (unspecialized) cells take on individual characteristics and reach their mature (specialized) form and function \cite{NCI_Cell_Differentiation})}, if one exists. To infer the dynamic process from static scRNA-seq data, many computational methods have been developed, including the methods that infer cell pseudotime (also known as trajectory inference) \cite{pseudotime}, RNA velocity \cite{rnavelocity}, and cell potency---three concepts that carry related but distinct meanings. \re{Specifically, pseudotime is a quantitative measure of progress through a biological process \cite{pseudotime}; RNA velocity is the time derivative of the gene expression state \cite{rnavelocity}; cell potency, also known as stemness, is a quantitative measure that indicates the differentiation capacity of a cell \cite{mushtaq2020cellstemness}.}\\

\noindent\textbf{Pseudotime vs. RNA velocity vs. potency}

\re{Although pseudotime, RNA velocity, and potency all aim to quantify dynamic cellular processes based on static scRNA-seq data, they provide distinct biological insights. Figure~\ref{fig:3d_compare} depicts the cellular differentiation process starting from colonic stem cells (CSCs) to differentiated cell types, including transit amplifying cells, cells under endoplasmic reticulum stress, goblet and deep crypt secretory (DCS) cells, and early or late enterocytes. With various computational methods applied to this differentiation dataset, distinct biological insights can be revealed by the outputted pseudotime, RNA velocity, and cell potency. As Figure~\ref{fig:3d_compare}b-d shows, pseudotime inference reveals both the lineage information (i.e., the brown, pink, green, and orange ``branches'') and numeric pseudotime values (represented by the numbers between 0 and 1 in each branch) that describe the relative position of each cell within its corresponding lineage. Unlike pseudotime, RNA velocities do not directly provide lineage information but show the local direction and speed of individual cell transitions using ``arrows" with varying lengths. While the overall pattern of these arrows may suggest lineages, the lineage information is much less straightforward compared to pseudotime. Last, cell potency does not indicate any multi-lineage information, as it is a numeric value for each cell (i.e., the ``height'' of the landscape), representing the cell's relative potency among all cells.}

\re{In addition to the differences in the biological information that pseudotime, RNA velocity, and potency can reveal, the computational methods to obtain such information also differ in the following four aspects. First, while the inference of cell pseudotime and RNA velocity is suitable for all cellular dynamic processes such as immune responses and development, cell potency estimation is specific to the cell differentiation process. 
Second, while pseudotime inference typically requires users to specify the starting point of a dynamic cellular process, RNA velocity inference and cell potency estimation do not have this requirement due to their inherent directional nature. RNA velocity indicates the direction of cell state change, and a higher potency value indicates a more stem-like cell. However, because of this difference, discrepancies may exist between pseudotime and the other two concepts. In Figure~\ref{fig:3d_compare}b-d, pseudotime inference uses the specification that CSC is the root cell type, while RNA velocities and cell potency indicate a direction from proCSC to CSC, which differs from the established knowledge. Third, the methods for estimating \re{pseudotime}, RNA velocity, and cell potency have distinct inputs and outputs (Table~\ref{input_output}). %The common input required by all three types of methods is a gene-by-cell expression count matrix processed from scRNA-seq data. 
\re{Pseudotime} inference methods (e.g., Monocle \cite{pseudotime} and Slingshot \cite{slingshot}) require users to input (1) cells' low-dimensional embeddings (e.g., embeddings by the principal component analysis (PCA), t-distributed stochastic neighbor embedding (t-SNE) \cite{tsne}, or uniform manifold approximation and projection (UMAP) \cite{mcinnes2018umap}) projected from a gene-by-cell expression matrix, (2) cells' cluster or type labels, and (3) the root cell of a trajectory (or multiple trajectories) based on users' prior knowledge. The output of pseudotime inference methods is a cell-by-trajectory pseudotime matrix where each entry marks a cell's pseudotime in a trajectory. RNA velocity inference methods (e.g., velocyto \cite{rnavelocity} and scVelo \cite{scvelo}) require two inputted gene-by-cell count matrices (of the same dimensions), which correspond to the genes' spliced and unspliced read counts, respectively, in the cells. The output of RNA velocity inference methods is a cell-by-gene velocity matrix, in which every cell has a velocity vector with a length equal to the number of genes. Typically, cells' velocity vectors are visualized together with the cells in a two-dimensional t-SNE or UMAP plot. Cell potency estimation methods, the focus of this article, all require a gene-by-cell expression matrix, and some methods also need a protein-protein interaction (PPI) network and/or a gene-gene similarity matrix. The output of cell potency estimation methods is a cell potency vector, whose length is equal to the number of cells. Figure~\ref{fig:3d_compare}b-d illustrates the different outputs of the three types of methods. In particular, pseudotime inference methods and cell potency estimation methods have the same output format when a single trajectory is specified for pseudotime inference; that is, each cell receives one pseudotime value and one potency value. However, it remains an open question whether any correspondence exists between cells' pseudotime values and potency values. Fourth, the three types of methods rely on different core algorithms. Most pseudotime inference methods use the minimum spanning tree algorithm to connect user-inputted cell clusters into a trajectory or trajectories. Most RNA velocity inference methods fit ordinary differential equations to estimate cells' velocity vectors. Unlike the first two types, computational methods that estimate cell potency employ diverse algorithms, ranging from summary statistic calculations to supervised learning algorithms.}\\

\noindent\textbf{Experimental and computational approaches to study cell potency}

\re{Knowing cell differentiation states helps researchers discover biological insights, such as the developmental relationships among various cell types \cite{cellTypeDevelopment, cellTypeDevelopment2}, the effects of a nanozyme-based eye drop on corneal epithelium regeneration \cite{tissueRenewal}, and changes in cell differentiation potential as mice age \cite{aging}. Therefore, both biological and computational researchers have employed various ways to identify pluripotent stem cells (PSCs) and quantify the potency of individual cells.}

\re{Experimentally, the approaches \cite{experiment, ref_lineage_tracing, ref_lineage_tracing2} to evaluate cell potency include (1) measuring the expression levels of the transcription factors that are known to play a key role in helping cells maintain pluripotent, (2) inducing PSCs in vitro and checking the cells' ability to differentiate into three germ layers (in vitro differentiation assay), (3) injecting test cells into various host environments (e.g., immunocompromised mice, blastocysts, tetraploid embryos, viable embryos, diploid or tetraploid blastocysts) and assessing the contribution of the injected cells to tissues from all three germ layers or the viability of the resulting chimera, and (4) conducting lineage-tracing experiments to track the progeny of individual cells over time.}

\re{While these experimental approaches provide direct and reliable ways for quantifying cell potency, they are costly and time-consuming. Therefore, leveraging scRNA-seq data, numerous computational methods have been developed to quantify individual cells' potency.} Ideally, cell potency estimation methods should assign the highest potency to stem cells and the lowest potency to fully differentiated cells. Accurate estimation of cell potency should reveal the hidden differentiation status of individual cells, thus expanding the capacity of scRNA-seq data analysis. \re{To the best of our knowledge, there are $14$ cell potency estimation methods, including CytoTRACE \cite{gulati2020cytotrace}, \re{ORIGINS \cite{senra2022origins}}, NCG \cite{ni2021accurate}, stemID \cite{grun2016stemID}, SLICE \cite{guo2017slice}, dpath \cite{gong2017dpath}, scEnergy \cite{jin2018scEnergy}, \re{cmEntropy \cite{kannan2021cmEntropy}}, SCENT \cite{teschendorff2017SCENT}, MCE \cite{shi2020mce}, 
SPIDE \cite{xu2022SPIDE}, mRNAsi \cite{malta2018machine}, CCAT \cite{teschendorff2021ccat}, \re{and FitDevo \cite{zhang2022fitdevo}}.} Based on the assumptions and approaches of these methods, we categorized them into three categories, including \re{three} \textit{\re{average-based} methods} (CytoTRACE, \re{ORIGINS},  and NCG), \re{eight} \textit{entropy-based methods} (stemID, SLICE, dpath, scEnergy, \re{cmEntropy}, SCENT, MCE, and SPIDE), and \re{three} \textit{correlation-based methods} (mRNAsi, CCAT, \re{and FitDevo}). As an overview, \re{average-based} methods summarize a subset of genes' expression levels into an average-like statistic to represent a cell's potency; entropy-based methods define an entropy-like measure to describe the expression-level distribution of genes or gene groups in a cell; and correlation-based methods use externally defined gene importance scores to calculate a correlation between the scores and each cell's gene expression levels. Since the single-cell field lacks a summary and comparison of cell potency methods, we are motivated to categorize the existing methods and help users understand them. \\

\noindent\textbf{This review}

Here, we review \re{$14$} published computational methods mentioned above for estimating cell potency from scRNA-seq data from multiple perspectives, including the definition, resolution, and calculation of cell potency, as well as the input data, including the external information such as gene ontology (GO) and a PPI network. We include only the methods that do not require users to have prior knowledge of the root cell, i.e., the starting point of cell differentiation, which is required for pseudotime inference. For this reason, we do not include \re{Palantir \cite{palantir}, a method that first infers pseudotime and then estimates cell potency from the inferred cell pseudotime values.} Tables~\ref{tab:summary} and \ref{intuition} summarize the \re{$14$} methods' characteristics and high-level intuitions, respectively.

During our detailed review of these methods, we spot \re{inconsistencies} between several methods' descriptions in their publications and their code implementations in the corresponding software packages/scripts, whose version numbers or latest update dates when accessed (if the version number is unavailable) are listed in Table~\ref{tab:summary}. We summarize these \re{inconsistencies} in \re{Table~\ref{tab:discrepancy}}. \re{We also list the number of parameters of the $14$ methods in Table~\ref{tab:total_parm} to reflect each method's complexity. More specifically, Table~\ref{tab:justify_tunable} summarizes the parameters tunable by users and the parameters with default values justified by empirical evidence in the methods' publications.} In Figure~\ref{fig:radar}, we use radar plots to compare the \re{$14$} methods' complexities in three dimensions: total number of parameters, number of inputs, and number of \re{inconsistencies}.

In the following sections, we will introduce mathematical notations to unify the notations used in the \re{$14$} methods, summarize these methods into the three categories: \re{average-based} methods, entropy-based methods, and correlation-based methods (illustrated in Figure~\ref{fig:cellpotency}), and provide a detailed explanation for each method.

\section*{Unified mathematical notations}
Before describing the \re{$14$} methods, we introduce the following mathematical notations to unify the notations used in different methods and facilitate our discussion.
\begin{itemize}
    \item $X \in \mathbb{Z}_{\ge 0}^{p \times n}$: a gene-by-cell count matrix, assuming we have $p$ genes and $n$ cells. The $(i,j)$-th entry $X_{ij}$ indicates the read or UMI (unique molecular identifier) count, which represents the measured expression level, of gene $i$ in cell $j$. The $j$-th column $X_{\cdot j}$ is a $p$-dimensional column vector consisting of all $p$ genes' counts in cell $j$.
    \item $S \in \mathbb{R}_{\ge 0}^{p \times n}$: a gene-by-cell normalized expression matrix, in which the cell library size is normalized out. The $(i,j)$-th entry $S_{ij}$ is the normalized expression level of gene $i$ in cell $j$. Note that the purpose of normalization is to make gene expression levels comparable across cells, and different methods may use different normalization approaches.

    \item $Z \in \mathbb{R}_{\ge 0}^{p \times n}$: a gene-by-cell log normalized expression matrix (i.e., applying log transformation on $S$). The $(i,j)$-th entry $Z_{ij}$ is the log normalized expression level of gene $i$ in cell $j$. Note that different methods may use different log-normalizing approaches. The $j$-th column $Z_{\cdot j}$ is a $p$-dimensional column vector consisting of all $p$ genes' log normalized expression levels in cell $j$.
    \item $A \in \{0,1\}^{p \times p}$: a gene-by-gene binary adjacency matrix from a user-inputted PPI network. $A$ is symmetric. For $i \neq i^\prime$, the $(i,i^\prime)$-th entry $A_{ii^\prime} = 1$ means that an edge exists between genes $i$ and $i^\prime$ in the PPI network, and $A_{ii^\prime} = 0$ means otherwise. Note that $A_{ii}=0$ in SCENT and SPIDE, but $A_{ii}=1$ in MCE.
    \item $N(i)$: a set of gene $i$'s neighboring genes in the PPI network (i.e., the genes that have edges connecting with gene $i$). The size of $N(i)$ is $\sum_{i^\prime=1}^p A_{ii^\prime}$.
    \item $\Pi \in [0,1]^{p \times n}$: a gene-by-cell steady-state probability matrix. The $(i,j)$-th entry $\Pi_{ij}$, defined based on $X_{ij}$, $S_{ij}$, or $Z_{ij}$ (different methods may have different definitions), represents the steady-state probability of gene $i$ in cell $j$. Intuitively, $\Pi_{ij}$ serves as a weight for gene $i$ in cell $j$. For SCENT and SPIDE, if we fix the total expression level of gene $i$'s neighboring genes in cell $j$, i.e., $\sum_{i^\prime \in N(i)} Z_{i^\prime j}$, then $\Pi_{ij}$ increases with $Z_{ij}$; if we fix $Z_{ij}$, then $\Pi_{ij}$ increases with $\sum_{i^\prime \in N(i)} Z_{i^\prime j}$. For MCE, $\Pi_{ij} = S_{ij} := X_{ij}/\left(\sum_{i^\prime = 1}^p X_{i^\prime j}\right)$ and thus increases with $S_{ij}$.
    \item $R_j \in [-1,1]^{p \times p}$: a gene-by-gene correlation matrix of cell $j$. The $(i,i^\prime)$-th entry $R_{ii^\prime j}$ is the Pearson correlation calculated based on \[\{(Z_{ij^\prime}, Z_{i^\prime j^\prime}): \text{cell } j^\prime \text{ is among the $K$ nearest neighbors of cell } j\}\,.\] Specifically, each cell $j$'s $K$ nearest neighbors are defined based on the Pearson correlation applied to $Z$, i.e., the distance between cells $j$ and $j^\prime$ is $1 - \text{cor}(Z_{\cdot j}\,{,} \, Z_{\cdot j^\prime})$.
    \item $P_j \in [0,1]^{p \times p}$: a gene-by-gene transition probability matrix of cell $j$. The $(i,i^\prime)$-th entry $P_{ii^\prime j}$ represents the transition probability from gene $i$ to gene $i^\prime \in N(i)$ in cell $j$. This transition probability is defined as an aggregation of the gene expression and PPI information ($A$), including $Z_{ij}$, $\{Z_{i^\prime j}: i^\prime \in N(i)\}$, as well as $R_{ii^\prime j}$. Different methods may aggregate these sources of information in different ways. 
    \item $W_j \in \mathbb{R}_{\ge 0}^{p \times p}$: a gene-by-gene edge weight matrix of cell $j$. The $(i,i^\prime)$-th entry $W_{ii^\prime j}$ represents the edge weight between gene $i$ and gene $i^\prime \in N(i)$. $W_{ii^\prime j}$ is defined based on $Z_{ij}$, $Z_{i^\prime j}$ and $R_{ii^\prime j}$, and different methods may have different definitions.

    \item $E_j$: the entropy for cell $j$ calculated by an entropy-based method. 
    \item  $\mathrm{\it{NE}}_j$: the normalized entropy for cell $j$ ($E_j$ is normalized by a constant $C$; the definition of $C$ differs across entropy-based methods).
\end{itemize}

\section*{Category 1: \re{average-based} methods}
Referred to as the \re{average-based} methods, CytoTRACE \cite{gulati2020cytotrace}, \re{ORIGINS \cite{senra2022origins}}, and NCG \cite{ni2021accurate} summarize the gene expression levels in a cell into an average-like summary statistics to represent the potency of the cell. \re{Among the three methods, CytoTRACE requires the fewest user inputs: only the gene-by-cell count matrix. In short, CytoTRACE first selects potency-related genes and then assigns a higher potency to the cells exhibiting higher expression levels of the selected genes. In addition to the gene expression data, ORIGINS requires a PPI network. Briefly, ORIGINS defines the potency of a cell as the sum of the cell-specific PPI edge weights, where each edge weight is the product of the expression levels of the two genes connected by the edge. Besides the gene expression data and the PPI network, 
NCG further relies on gene-gene similarities pre-calculated based on genes' GO terms. First, NCG assigns weights to genes based on the PPI network and gene-gene similarities. Then, NCG defines the potency of a cell as the weighted sum of gene expression levels in the cell.}

\subsection*{CytoTRACE}

CytoTRACE \cite{gulati2020cytotrace} relies on the assumption that a cell's \textit{gene count} (i.e., the number of expressed genes) reflects the cell's differentiation status. Accordingly, CytoTRACE first selects the genes whose expression levels highly correlate with the gene count across cells. Then CytoTRACE defines the \textit{gene count signature (GCS)} as the geometric mean of the selected genes' expression levels in each cell. Finally, CytoTRACE performs smoothing on the cells' GCSs and then converts the smoothed GCSs into ranks, which serve as the final estimates of the cells' potency.

Technically, CytoTRACE first transforms $X \in \mathbb{Z}_{\ge 0}^{p \times n}$ to CPM $S \in \mathbb{R}_{\ge 0}^{p \times n}$ (Count Per Million; $S_{ij} = X_{ij} / \sum_{i^\prime=1}^p X_{ij} \times 10^6$) and then to $Z \in \mathbb{R}_{\ge 0}^{p \times n}$, with $Z_{ij} = \log_2\left(
C_j \frac{S_{ij}}{\sum_{i^\prime=1}^p S_{i^\prime j}}+1 \right)$, where $C_j = \sum_{i^\prime=1}^p \mathbb{I}(X_{i^\prime j} > 0)$ is cell $j$'s gene count. Second\footnote{The number of cells in the code implementation differs from what is described in the publication. Instead of $n$ cells, it becomes $\tilde{n}$ selected cells, in which at least one of the $1{,}000$ genes selected in the sub-step $1$ of the fourth step (the step that filters genes) is expressed. However, in the manuscript, no filtering of cells is mentioned.}, CytoTRACE calculates the Pearson correlation between all $n$ cells' gene counts $(C_1, C_2, \ldots, C_n)\tran$ and gene $i$'s log normalized expression levels in the $n$ cells $(Z_{i1}, Z_{i2}, \ldots, Z_{in})\tran$, for each gene $i =1, \ldots, p$. Third, CytoTRACE selects the top $200$ genes that have the highest correlation values to compute the GCS for each cell, obtaining $G=(G_1, G_2, \ldots, G_n)\tran \in \mathbb{R}_{\ge 0}^{n}$, where $G_j$ is cell $j$'s GCS, defined as the geometric mean of the $200$ genes' log normalized expression levels in cell $j$. However, the GCS is defined as the arithmetic mean instead in the code implementation, different from the publication (for the detailed explanation, see Supplementary Material). Fourth, CytoTRACE performs smoothing on $G$ by taking the following sub-steps.
\begin{enumerate}
    \item Filter the rows of $Z$ to retain only the genes that are expressed in at least $5\%$ of the $n$ cells. For each retained gene $i$, compute a \textit{dispersion index} as the gene's variance over its mean across the $n$ cells, i.e., $\frac{1}{n-1}\sum_{j=1}^n (Z_{ij} - \bar{Z}_{i})^2/\bar{Z}_{i}$, where $\bar{Z}_{i} = \frac{1}{n}\sum_{j=1}^n Z_{ij}$. Further, filter the rows of $Z$ to retain only the $1{,}000$ genes with the largest dispersion indices. Hence, the dimension of $Z$ becomes $1{,}000 \times n$.
    
    We note that this gene selection sub-step is performed before the second step above in the code implementation, and the selected $1{,}000$ genes are used to select $\tilde{n}$ cells (see footnote), creating an inconsistency between the code implementation and the publication. Specifically, in the second step above, each Pearson correlation is calculated across all $n$ cells in the publication but only the $\tilde{n}$ cells in the code implementation. Hence,  the top $200$ genes selected in the third step above differ between the code implementation and the publication. Also, the dimension of $Z$ in the code becomes $1{,}000 \times \tilde{n}$. Thus, in the following sub-steps, the cell number $n$ becomes $\tilde{n}$ in the code.
    %Notably, in CytoTRACE's code implementation, the dimension of $Z$ is different from what is described in the publication. Instead of $1{,}000 \times n$, the dimension in the code is $p \times \tilde{n}$, with all $p$ genes and the selected $\tilde{n} < n$ cells. The $\tilde{n}$ cells are selected as those with at least one of the $1{,}000$ expressed. Hence, in the following steps, the cell number $n$ is replaced by $\tilde{n}$ in the code.
    \item Obtain a cell-by-cell Markov matrix $B \in [-1, 1]^{n \times n}$ in two steps.
    \begin{enumerate}
        \item Compute a cell-by-cell similarity matrix $D \in [-1, 1]^{n \times n}$, in which $D_{jj^\prime}$ is the Pearson correlation between cell $j$ and cell $j^\prime$ based on the retained $1{,}000$ genes' expression levels in $Z$, i.e., the Pearson correlation between the two columns $Z_{\cdot j}$ and $Z_{\cdot j^\prime}$.
        \item Transform $D$ to $B$ as
        \[
        B_{jj^\prime} = \left\{ \begin{array}{ll}
          0   &  \text{if } j = j^\prime \text{ or } D_{jj^\prime} < \tau\\
          D_{jj^\prime}   & \text{otherwise}
        \end{array}\right.\,,
        \]
        where $\tau = \frac{1}{n^2}\sum_{j=1}^n\sum_{j^\prime=1}^n D_{jj^\prime}$ is the mean of $D$.
    \end{enumerate}
    \item Transform $G$ to $G_0$ using non-negative least squares regression, which finds an $n$-dimensional vector $\beta_0 = \argmin _{\beta \geq 0}\|G - B\beta\|_{2}^2$, and defining $G_0 = B\beta_0$.
    \item Simulate a diffusion process for $T=10{,}000$ iterations or until convergence:
\begin{equation*}
    G_{t+1} = \alpha \cdot B G_{t}+(1-\alpha) \cdot G_{0}\,;\;t=0,\ldots,T-1\,,
\end{equation*}
where the degree of diffusion $\alpha = 0.9$.    
\end{enumerate}

Finally, CytoTRACE estimates the \textit{potency} by converting 
the $n$ values in $G_{T}$ from the last iteration into ranks $n$ to $1$ (with $n$ indicating the maximum value in $G_{T}$). 

The intuition behind CytoTRACE is that a cell has a higher potency if it has more genes expressed.

 Compared with the other $13$ methods, CytoTRACE has one of the most complicated procedures.  The parameters in CytoTRACE include the two gene numbers ($200$ and $1{,}000$) used to calculate $G$ and $D$, respectively, the percentage threshold used to filter out genes of $5\%$, as well as the degree of diffusion $\alpha$ in the diffusion process. In total, we count CytoTRACE to have four parameters, \re{among which the value choices of the gene number $200$ and $\alpha = 0.9$ are justified by empirical evidence in the CytoTRACE publication \cite{gulati2020cytotrace}. 
 In the \texttt{CytoTRACE} R package (v 0.3.3), none of these four parameters are tunable by users.}

\subsection*{ORIGINS}
\re{ORIGINS \cite{senra2022origins} defines cell potency from a gene-by-cell count matrix $X \in \mathbb{Z}_{\ge 0}^{p \times n}$ and a PPI network $A \in \{0,1\}^{p \times p}$.  ORIGINS calculates a weighted adjacency matrix $W_j \in \mathbb{R}_{\ge 0}^{p \times p}$ for each cell 
$j$, with the $(i, i^\prime)$-th entry set as $W_{ii^\prime j} = A_{ii^\prime}X_{ij}X_{i^\prime j}$. Then, ORIGINS sums over the off-diagonal entries of $W_j$ to obtain $L_j = \sum_{i = 1}^p \sum_{i^\prime \neq i} W_{ii^\prime j}$, which is assumed to represent cell $j$'s activity level associated with the differentiation process. ORIGINS scales $L_j$ to obtain the \textit{potency} $L_j^* \in [0,1]$ for cell $j$ as 
\begin{equation*}
    L_j^* =\frac{L_j - \min_{j^{\prime \prime} \in \{1, \cdots, n\}}L_{j^{\prime \prime}}}{\max_{j^\prime \in \{1, \cdots, n\}}L_{j^\prime} - \min_{j^{\prime \prime} \in \{1, \cdots, n\}}L_{j^{\prime \prime}}}\,.
\end{equation*}}

\re{The intuition behind ORIGINS is that a cell has a higher potency if the genes connected in the PPI network are more highly expressed.}

\re{We count ORIGINS to have zero parameters.}

\subsection*{NCG}

NCG \cite{ni2021accurate} defines cell potency from scRNA-seq data, a PPI network, and GO annotations\footnote{Note that the required inputs of NCG include a count matrix, a PPI network, and a gene-gene similarity defined based on GO annotations.}. First, NCG performs quantile normalization on the inputted count matrix $X = (X_{ij})$ to obtain $S = (S_{ij})$. Then, NCG log transforms $S$ to obtain $Z = (Z_{ij})$. In the publication, if $S_{ij} < 1$, $Z_{ij} = \log_2(S_{ij} + 1.1)$; otherwise, $Z_{ij} = \log_2(S_{ij} + 0.1)$ (Noted that in NCG's code implementation,  $Z_{ij}$ is always $\log_2(S_{ij} + 0.1)$; if $S_{ij} < 1$, then $S_{ij}$ is set to $1$, and $Z_{ij}$ becomes a constant $\log_2 1.1$, which we find unreasonable). Next, NCG defines a \textit{weighted edge clustering coefficient of the gene (ECG)} for each gene in both the scRNA-seq data and the PPI network. The ECG for gene $i$ is $\mathrm{\it{ECG}}_i = \sum_{i^\prime=1}^{p} w_{ii^\prime} \times \mathrm{\it{ECC}}_{ii^\prime}$, in which $p$ is the number of genes in both the scRNA-seq data and the PPI network, $w_{ii^\prime}$ is the similarity between gene $i$ and gene $i^\prime$ defined as the Kappa statistic applied to the two genes' GO annotations (detailed below), and $\mathrm{\it{ECC}}_{ii^\prime}$ is the \textit{edge clustering coefficient (ECC)} (defined as $|N(i) \cap N(i^\prime)|/(\min(N(i), N(i^\prime))-1)$, the number of common neighbors between gene $i$ and gene $i^\prime$ divide by the minimum degree of the two genes minus $1$; it is set to $0$ when the denominator is $0$). Note that the ECC measures the degree of closeness of two genes in a PPI network. Intuitively, if a gene has a higher ECG value, it is similar to more genes based on the PPI network and GO annotations. 

Next, NCG computes the \textit{potency} for cell $j$ as \[\mathrm{\it{NCG}}_j = \frac{1}{\max_{j^\prime \in \{1, \cdots , n\}}\mathrm{\it{NCG}}_{j^\prime}}\sum_{i=1}^{p} Z_{ij} \times \mathrm{\it{ECG}}_i \,.\] A high value of $\mathrm{\it{NCG}}_j$ indicates a high potency value of cell $j$.

Since the Kappa statistic for each gene pair is a required input that users must specify, the NCG paper \cite{ni2021accurate} does not detail the calculation of the Kappa statistic. Based on literature \cite{kappa,huang2007david_kappa}, the Kappa statistic $w_{ii^\prime}$ between gene $i$ and gene $i^\prime$ is defined as follows. Given a total of $N$ GO terms, we could construct the following confusion matrix based on the two genes' GO terms.

\begin{table}[h]
    \centering
\begin{tabular}{ cc|c|c|c } 
  \multicolumn{5}{c}{Gene $i^\prime$} \\ 
\multirow{5}{*}{\rotatebox{90}{Gene $i$}} & & $1$ & $0$ & Row Total \\ \cline{2-5}
    & $1$ &$N_{ii^\prime}^{11}$ & $N_{ii^\prime}^{10}$ &$N_{i\cdot}^{1*}$\\ \cline{2-5}
    & $0$&$N_{ii^\prime}^{01}$ &$N_{ii^\prime}^{00}$ &$N_{i\cdot}^{0*}$\\ \cline{2-5}
    & Column Total &$N_{\cdot i^\prime}^{*1}$ &$N_{\cdot i^\prime}^{*0}$ & ${N}$\\
\end{tabular}
\end{table}
In the confusion matrix, we divide the $N$ GO terms based on whether they belong to gene $i$ and/or gene $i^\prime$, with the row or column label of $1$ indicating belonging and $0$ otherwise. Then for example, $N_{i i^\prime}^{11}$ indicates the number of GO terms shared by genes $i$ and $i^\prime$, and $N_{i i^\prime}^{10}$ means the number of GO terms belonging to gene $i$ but not to gene $i^\prime$. The Kappa statistic is then defined as $w_{ii^\prime} = \frac{O_{ii^\prime} - E_{ii^\prime}}{1 - E_{ii^\prime}}$ where $O_{ii^\prime} = \frac{N_{ii^\prime}^{11}+N_{ii^\prime}^{00}}{N}$ and $E_{ii^\prime} = \frac{N_{i\cdot}^{1*}N_{i\cdot}^{*1} + N_{\cdot i^\prime}^{0*}N_{\cdot i^\prime}^{*0}}{N^2}$.

The intuition behind NCG is that, with a subset of genes identified as closely connected in the PPI network and sharing similar GO terms, a cell has a higher potency if the identified genes have higher expression in it.

The parameters in NCG include the offset $1.1$ used in the log transformation ($Z_{ij} = \log_2(S_{ij} + \text{offset})$) and the cutoff value of $1$ for $S_{ij}$ to decide which offset value ($1.1$ or $0.1$) to use. We do not count the other offset $0.1$ used for $S_{ij} \ge 1$ as a parameter because we interpret this offset value to make $Z_{ij}$ continuous at $S_{ij} = 1$ in the code implementation (however, this interpretation does not hold for the publication, which differs from the code implementation). In total, we count NCG to have two parameters. \re{Neither of the two parameters' default value choices was justified by empirical evidence, and neither parameter is tunable by users in NCG's R script (update on September 1, 2021, when accessed).}

\section*{Category 2: entropy-based methods}
Out of the \re{$14$} methods, \re{eight} methods (stemID \cite{grun2016stemID}, SLICE \cite{guo2017slice}, dpath \cite{gong2017dpath}, scEnergy \cite{jin2018scEnergy}, \re{cmEntropy \cite{kannan2021cmEntropy},} SCENT \cite{teschendorff2017SCENT}, MCE \cite{shi2020mce}, and SPIDE \cite{xu2022SPIDE}) use various ways to define and calculate the entropy, i.e., the cell potency measure, of each cell or cell cluster. These methods are based on the rationale that stem cells should have higher entropy values (i.e., \re{genes have similar expression levels in stem cells}), while mature cells should have lower entropy values (i.e., \re{genes exhibit different expression levels in mature cells.}). 

Based on the methodological details, we further separate the seven methods into three sub-categories, depending on the resolution of cell potency (at the cell-cluster or single-cell level) and the use of a PPI network. The first sub-category includes stemID, the only method that calculates potency at the cell-cluster level. stemID first finds cell clusters and then uses clusters' similarities and relative gene expression levels to calculate entropy for each cluster. The second sub-category includes SLICE, dpath, scEnergy, \re{and cmEntropy}, which use the relative or binarized expression levels of either gene groups \re{(SLICE and dpath)} or individual genes \re{(scEnergy and cmEntropy)} to define the potency of each cell. The third sub-category requires a PPI network and includes SCENT, MCE, and SPIDE, which use different ways to define genes' steady-state probabilities and gene-to-gene transition probabilities based on a PPI network and gene expression levels. Then, they aggregate these probabilities into an entropy-like measure for each cell. Finally, they normalize the measure as a potency estimate for each cell.

\subsection*{stemID}

Among the seven entropy-based methods, stemID \cite{grun2016stemID} is the only one that estimates the potency of cell clusters instead of cells. Hence, stemID requires users to run the clustering method RaceID2 \cite{grun2016stemID}, which was developed by the same authors. The RaceID2 result contains each cell's cluster label and each cluster's medoid. 

First, stemID links all pairs of cluster medoids and determines the significance of each link as follows: every cell is projected to the closest link
(by vertical distance in the embedding space by multi-dimensional scaling, which preserves the pairwise cell distances\footnote{The distance between cells $j$ and $j'$ is defined as one minus the Pearson correlation of the two cells' count vectors $X_{\cdot j}$ and $X_{\cdot j'}$. The dimension of the multi-dimensional scaling embedding space is the number of positive eigenvalues of the cell distance matrix.}) among the links that connect the cell's cluster medoid to other cluster medoids; after all cells are projected, each link has its projected cells, whose number would then be converted to a p-value (considered significant if under a default significance level of $0.01$). Given the significant links, stemID computes the number of significant links involving each cluster $k$'s medoid and denotes the number as $l_k$. 

Next, stemID computes the \textit{entropy value} of cell $j$ as $E_j = -\sum_{i=1}^p S_{ij} \log_p S_{ij}$ (note that the negative sign was missing in the original article as a typo), where $S_{ij} := X_{ij}/\sum_{i^\prime=1}^p X_{i^\prime j}$ is the proportion of gene $i$'s expression level in cell $j$. 

Finally, stemID computes a \textit{stemness score} $s_k$ as the estimated \textit{potency} for each cell cluster $k$ as 
\[s_k = l_k \times \Delta E_k\,,\]
where $\Delta E_k$ is the difference between cluster $k$'s median cell entropy value (i.e., the median of the entropy values of the cells in cluster $k$) and the minimum of all clusters' median cell entropy values.

The intuition behind stemID is that a cluster has a higher potency if it has more clusters in proximity
and contains more cells with \re{genes exhibiting similar expression levels.}.

The parameter in stemID is the significance level of $0.01$ used for identifying significant links. In total, we count stemID to have one parameter. \re{This parameter's default value choice was not justified by empirical evidence, but the parameter is tunable by users in stemID's R script (update on March 3, 2017, when accessed).}
\\\\
Unlike stemID, the other \re{seven} entropy-based methods estimate the potency of each cell. 

\subsection*{SLICE}

SLICE \cite{guo2017slice} defines the potency measure, called scEntropy, for each cell based on the inputted gene expression matrix (which can be \re{Seurat's normalized data} or be on the scale of FPKM (Fragments Per Kilobase of transcript per Million mapped reads), RPKM (Reads Per Kilobase per Million mapped reads), or TPM (Transcripts Per Million)) and GO annotations. Specifically, SLICE calculates a cell-specific entropy based on gene groups instead of genes, unlike stemID; SLICE clusters genes into groups by performing $k$-means clustering on a gene-gene dissimilarity matrix, which the gene-gene dissimilarities are measured by one minus the Kappa statistic applied to the genes' GO terms (same as the Kappa statistic used in NCG), and this Kappa statistic matrix is a required user input. SLICE's R package provides the similarity matrices measured by Kappa statistics for human genes and mouse genes, respectively.  

Note that the default use of the $k$-means clustering might be problematic because the $k$-means clustering should be a gene-by-feature matrix instead of SLICE's symmetric gene-gene dissimilarity matrix.  

In detail, for each cell $j$, SLICE computes the scEntropy, which is defined as the \textit{potency} for cell $j$, as
\[ E_j = \frac{1}{B}\sum_{b=1}^B \sum_{m=1}^{M} - p_j^m (G_b,F_b) \log p_j^m (G_b,F_b)\,, 
\]
where $G_b$ is a random subsample of $p_s = 1{,}000$ genes with expression levels greater than $1$ in at least one cell (note that the original article incorrectly referred to $G_b$ as a ``bootstrap sample"), $F_b$ denotes the $M = \sqrt{p_s/2} = 22$ gene groups clustered from the genes in $G_b$, and $p_j^m (G_b,F_b) = \frac{c_j^m(G_b, F_b) + \alpha^m}{C_j(G_b, F_b) + A}$, in which $c_j^m(G_b, F_b)$ denotes the number of genes in $G_b$ expressed in cell $j$ and belonging to the $m$-th group in $F_b$, $\alpha^m$ denotes the proportion of genes in $G_b$ belonging to the $m$-th group in $F_b$ (i.e. the number of genes in $G_b$ belonging to he $m$-th group in $F_b$ divided by $p_s$), $C_j(G_b, F_b) = \sum_{m=1}^{M}c_j^m(G_b, F_b)$, and $A = \sum_{m=1}^{M} \alpha^m$. Hence, $p_j^m (G_b,F_b)$ measures the relative degree of activeness (i.e., proportion) of the $m$-th gene group in $F_b$ in cell $j$, and $E_j$ denotes cell $j$'s average entropy calculated on the proportions of $M$ gene groups across $B = 100$ subsamples. Here, the values of $M$ and $B$ are from SLICE's tutorials \cite{slice_demo}.

Note that SLICE treats gene expression counts as binary because it only considers if a gene is expressed or not in a cell in its scEntropy calculation. 

The intuition behind SLICE is that a cell has a higher potency if \re{the $M$ gene groups have similar proportions of expressed genes. However, the 
$M$ gene groups are defined within each gene subsample and are not consistent across different gene subsamples.}

The parameters in SLICE include the expression level threshold of $1$ used to filter out genes, the number of genes in each subsample ($p_s$), the number of gene groups ($M$), and the number of subsamples ($B$). In total, we count SLICE to have four parameters. \re{The default value choices of these four parameters were not justified by empirical evidence, but all four parameters are tunable by users in the \texttt{SLICE} R package (v 0.99.0).}

\subsection*{dpath}

dpath \cite{gong2017dpath} is similar to SLICE but uses a different way to partition genes into groups: dpath first log-transforms the inputted gene-by-cell TPM matrix to obtain $Z$ ($Z = \log (\text{TPM} + 
 1)$). Then, dpath performs a weighted non-negative factorization (NMF) on $Z$ to cluster genes into $M$ (default $5$) groups (called ``metagenes"). The factorization is in the form of  $Z \approx UV$, where $U$ is a gene-by-metagene non-negative matrix, and $V$ is a metagene-by-cell non-negative matrix satisfying $\sum_{m=1}^{M} V_{mj} = 1$ for all cell $j$ (i.e., every column of $V$ sums up to $1$, so $V_{mj}$ indicates the proportion of metagene $m$ in cell $j$). The dpath authors denoted $\mu = (\mu_{ij}) =UV$ as the expected gene-by-cell logTPM matrix. Given the observed gene-by-cell logTPM matrix $Z$, the dpath authors wrote the log-likelihood of $U$, $V$, and $\pi$ (a gene-by-cell weight matrix as a parameter to be estimated) as 
\begin{align*}
\ell(U, V, \pi) = \sum_{i=1}^p \sum_{j=1}^n \left[\pi_{ij} \cdot \log \mathrm{Poisson}(Z_{ij}~|~\mu_{ij}) \right. \left. + ~ (1 - \pi_{ij}) \cdot \log \mathrm{Poisson}(Z_{ij}~|~0.1)\right]\,,
\end{align*}
where $\mathrm{Poisson}( \cdot | \lambda)$ represents the probability mass function of the Poisson distribution with mean $\lambda$; $\pi_{ij}$ is the weight for gene $i$ in cell $j$; $\mu_{ij}$ is the expected logTPM of gene $i$ in cell $j$; and $0.1$ is the mean parameter of the Poisson distribution modeling the dropout events. However, we did not find it appropriate to use the Poisson distribution to model logTPM data. 

Then dpath estimates $U$, $V$, and $\pi$ through an iterative optimization process to maximize the log-likelihood $\ell(U, V, \pi)$, and this process is referred to as the ``weighted NMF.'' In detail, dpath performs two rounds of interations in the optimization process. First, dpath iteratively solves $U$ and $V$ by fixing $\pi$ across all iterations: $\pi_{ij} = 1$ if $Z_{ij} > 0$, and $\pi_{ij} = 0.1$ otherwise (note that in the dpath package, $\pi_{ij}$ was set to $0$ instead of $0.1$, the value written in the dpath paper); hence, non-zero $Z_{ij}$'s have larger weights than zero $Z_{ij}$'s. Second, dpath uses the $U$ and $V$ found by the first round of iterations as the initialization and iteratively solves $U$, $V$ and $\pi$; in the $t$-th iteration of the second round, $\pi_{ij}$ is updated as $\pi_{ij}^t = \frac{\mathrm{Poisson}(Z_{ij}~|~\mu_{ij}^t)}{\mathrm{Poisson}(Z_{ij}~|~\mu_{ij}^t) ~ + ~ \mathrm{Poisson}(Z_{ij}~|~0.1)}$, where $\mu_{ij}^t = \sum_{m=1}^{M}U_{im}^tV_{mj}^t$. 

Using $V$ outputted by the second round, dpath estimates the \textit{potency} for cell $j$ as 
\[E_j = - \sum_{m=1}^M V_{mj} \log V_{m
j}\,.\] 

The intuition behind dpath is that a cell has a higher potency if \re{its metagenes have similar expression levels}. 

The parameters in dpath include the number of metagenes ($M$), the mean parameter for the Poisson distribution modeling the dropout event (default parameter value $0.1$), and the value of $\pi_{ij} = 0.1$ used in the first round of the weighted NMF for $Z_{ij} = 0$. In total, we count dpath to have three parameters. \re{The default value choices of these three parameters were not justified by empirical evidence. In the \texttt{dpath} R package (v 2.0.1), users can tune the number of metagenes and the mean parameter for the Poisson distribution modeling the dropout event, but the value of $\pi_{ij} = 0.1$ when $Z_{ij} = 0$ is not tunable.}

\subsection*{scEnergy}
scEnergy \cite{jin2018scEnergy} is different from SLICE and dpath because it does not calculate the cell-specific entropy based on gene groups. Instead, scEnergy considers every gene and the neighboring genes in a gene network constructed from the inputted gene-by-cell expression matrix (which can be on the scale of TPM, FPKM, or UMI count). Specifically, scEnergy first constructs a gene network by connecting every two genes that have absolute Spearman correlations greater than a threshold (default $0.1$); the Spearman correlation is defined between two genes' log-transformed counts, \re{i.e., between $(S^\ast_{i1}, \ldots, S^\ast_{in})\tran$ for gene $i$ and $(S^\ast_{i^\prime 1}, \ldots, S^\ast_{i^\prime n})\tran$ for gene $i^\prime$, where $S^\ast_{ij} = \log_2 (X_{ij} + 1)$}. Then for cell $j$, scEnergy defines the \textit{energy} as
 \[E_j = - \sum_{i=1}^p Z^\ast_{ij} \log \frac{Z^\ast_{ij}}{\sum_{i^\prime \in N(i)}{Z^\ast_{i^\prime j}}}\,,\]
 where \re{$Z^\ast_{ij} = \frac{S^\ast_{ij} - \min_{i^{\prime \prime} \in \{1,\cdots,p\}}S^\ast_{i^{\prime \prime} j}}{\max_{i^\prime \in \{1,\cdots,p\}}S^\ast_{i^\prime j} - \min_{i^{\prime \prime} \in \{1,\cdots,p\}}S^\ast_{i^{\prime \prime} j}}$\,. } 
 
 Finally, scEnergy estimates the \textit{potency} for cell $j$ by re-scaling the energy $E_j$ as 
 \[\mathrm{\it NE}_j = \frac{(E_j/\overline{E})^2}{1 + (E_j/\overline{E})^2}\,,
 \]
 where $\overline{E}$ is the average energy of all cells, and $\mathrm{\it NE}_j$ is outputted as the estimated \textit{potency} of cell $j$. However, we were unclear about the intuition behind the $E_j$ definition (which was claimed to be based on maximum entropy and statistical thermodynamics, but we could not find the exact connection) and the re-scaling, whose rationale was not mentioned in the scEnergy publication \cite{jin2018scEnergy}. 

The rough intuition behind scEnergy is that a cell has a higher potency if \re{genes have similar expression levels in it}.

The parameter in scEnergy is the threshold of $0.1$ for the absolute Spearman correlations of gene pairs when constructing the gene network. In total, we count scEnergy to have one parameter. \re{The default value choice for this parameter was not justified by empirical evidence, but this parameter is tunable by users in scEnergy's MATLAB script (update on February 16, 2021).}

\subsection*{cmEntropy}

\re{cmEntropy \cite{kannan2021cmEntropy} was proposed to quantify the cardiomyocyte maturation status at the single-cell level using the Shannon entropy. The cmEntropy authors also found cmEntropy applicable to other tissues for quantifying cell potency. Given the inputted gene-by-cell count matrix $X$, cmEntropy defines the \textit{potency} for cell $j$ as 
\begin{equation*}
    E_j = - \sum_{i = 1}^p S_{ij} \log S_{ij}\,,
\end{equation*}
where $S_{ij} = X_{ij} / \left(\sum_{i^\prime=1}^p X_{i^\prime j}\right)\,$.}

\re{The intuition behind cmEntropy is that a cell has higher potency if its genes have more similar expression levels.}

\re{We count cmEntropy to have zero parameters.}

\subsection*{SCENT}

SCENT \cite{teschendorff2017SCENT} differs from stemID, SLICE, dpath, and scEnergy in the probabilities used for entropy calculation. Using the signaling entropy definition in the SCENT authors' previous work \cite{sr1, sr2}, SCENT takes as inputs (1) a log normalized gene-by-cell expression matrix $Z$, with the log normalized expression of gene $i$ in cell $j$ as $Z_{ij} = \log_2(S_{ij}+1.1)$, where $S_{ij}$ is $X_{ij}$ (the count of gene $i$ in cell $j$) multiplied by cell $j$'s specific scaling factor calculated externally and (2) a PPI network that specifies the gene-by-gene binary adjacency matrix $A \in \{0,1\}^{p \times p}$. Then for each cell $j$, SCENT calculates the \textit{signaling entropy} as 

\[E_j = -\sum_{i=1}^p \Pi_{ij} \sum_{i^\prime \in N(i)} P_{ii^\prime j} \log P_{ii^\prime j}\,.
\]

The $E_j$ formula has two key components. First, the steady-state probability of gene $i$ in cell $j$ is 
\[\Pi_{ij} = \frac{Z_{ij}  (A \cdot Z_{\cdot j})_i}{Z_{\cdot j}^\intercal  \cdot  (A  \cdot  Z_{\cdot j})}\,,
\]
where $Z_{\cdot j} \in \mathbb{R}_{\ge 0}^p$ is the log normalized gene expression vector of cell $j$, and $(A \cdot Z_{\cdot j})_i$ is the $i$-th element of the vector $(A \cdot Z_{\cdot j}) \in \mathbb{R}_{\ge 0}^p$ and is equivalent to $\sum_{i^\prime \in N(i)}Z_{i^\prime j}$. Second, the transition probability from gene $i$ to gene $i^\prime$ in cell $j$ is 
\[P_{ii^\prime j} = \frac{W_{ii^\prime j}}{\sum_{i'' \in N(i)}W_{ii''j}}\,,
\]
where $W_{ii^\prime j} = Z_{ij}Z_{i^\prime j}$, so $P_{ii^\prime j} = \frac{Z_{i^\prime j}}{\sum_{i'' \in N(i)}Z_{i''j}}$ is irrelevant to $Z_{ij}$ and only uses gene $i$'s neighbors in the PPI network. 

SCENT estimates the \textit{potency} for each cell $j$ as  
\[\mathrm{\it{NE}}_j = E_j / C\,,
\]
in which the normalization constant $C$ is the maximum signaling entropy defined as 
\[ C = \max_{P = (P_{ii^\prime})} \left(-\sum_{i=1}^p \Pi_{ij} \sum_{i^\prime \in N(i)} P_{ii^\prime} \log P_{ii^\prime}\right) = -\sum_{i=1}^p \Pi_{ij} \sum_{i^\prime \in N(i)} \left(\frac{A_{ii^\prime}\nu_{i^\prime}}{\lambda \nu_i}\right) \log \left(\frac{A_{ii^\prime}\nu_{i^\prime}}{\lambda \nu_i}\right) = \log \lambda\,,
\]
where $\lambda$ is the largest eigenvalue of the adjacency matrix $A$, and $\nu_i$ is the $i$-th element in the eigenvector $\nu$ corresponding to the largest eigenvalue. The last equality $C = \log \lambda$ was proven in \cite{loglambda}.

As the number of cells increases, SCENT becomes computationally intensive. Therefore, CCAT \cite{teschendorff2021ccat}, a faster proxy of SCENT, was proposed. Since CCAT's approach does not involve entropy calculation, it belongs to another category we define as ``correlation-based methods," and we will discuss CCAT in the next section. \re{Later, the same research group proposed LandSCENT \cite{LandSCENT}, a direct extension of SCENT. However, since the step in LandSCENT that computes cell potency is identical to that of SCENT, we have not included LandSCENT in this review.}  

The intuition behind SCENT is that a cell $j$ has a higher potency if its ``influential" genes (i.e., gene $i$ is influential in cell $j$ if it has a large $\Pi_{ij}$, indicating gene $i$ is highly expressed in cell $j$ or has neighboring genes highly expressed in cell $j$) have \re{neighboring genes expressed at similar levels}.

The parameter in SCENT is the offset value of $1.1$ used in log transformation $Z_{ij} = \log_2(S_{ij} + \text{offset})$. In total, we count SCENT to have one parameter. \re{The default value choice of this parameter was not justified by empirical evidence, but this parameter is tunable by users as long as the value is greater than $1$ to ensure $Z_{ij}$ is positive. Users will input the $Z$ matrix, instead of the $S$ matrix, into the SCENT package.}
\\\\
The last two entropy-based methods, MCE \cite{shi2020mce} and SPIDE \cite{xu2022SPIDE}, are similar to SCENT in that they all use genes' expression levels and genes' relationships in a PPI network to compute an entropy for each cell. 

\subsection*{MCE}

MCE (Markov Chain Entropy) \cite{shi2020mce} defines the \textit{entropy} for each cell $j$ as 
\[E_j = \max_{P_j = (P_{ii^\prime j}) \text{ s.t. } P_j \Pi_{\cdot j} = \Pi_{\cdot j}}\left(- \sum_{(i,i^\prime)\in \hat{G}} \Pi_{ij}{P_{ii^\prime j}}\log(\Pi_{ij}P_{ii^\prime j})\right)\,,\]
where $(i,i^\prime)$ indicates an edge between genes $i$ and $i^\prime$ in a PPI network, and $\hat{G}$ is a set including all edges in the PPI network, including all self-loop edges. In the above definition of $E_j$, $\Pi_{ij} = X_{ij}/ (\sum_{i^\prime=1}^p X_{i^\prime j})$ is the steady-state probability of gene $i$ in cell $j$, and $\Pi_{\cdot j}$ is cell $j$'s steady-state probability vector; $P_{ii^\prime j}$ is the $(i,i^\prime)$-th element in cell $j$'s transition matrix $P_j$ that satisfies $P_j \Pi_{\cdot j} = \Pi_{\cdot j}$ and $P_j \mathbf{1} = \mathbf{1}$. The computation of $E_j$ is solved as a convex optimization problem. 

MCE's final estimate of cell $j$'s \textit{potency} is $\mathrm{\it{NE}}_j = E_j/C$, 
 in which the normalization constant $C = \log (\sum_{i=1}^p d_i)$, with $d_i$ representing the degree of gene $i$ in the PPI network. 
 
 Two major differences between MCE and SCENT are as follows. First, the inputted PPI network for MCE can have all ones on the diagonal of the adjacency matrix, while for SCENT the diagonal must have all zeros. Second, the transition probability matrix $P_j$ of cell $j$ is optimized given the steady-state probability $\Pi_j$ for MCE, while $P_j$ is computed based on the log normalized gene expression levels for SCENT. 

The intuition behind MCE is that a cell has a higher potency if its weighted gene-to-gene transition probabilities are \re{similar}.

We count MCE to have zero parameters.

\subsection*{SPIDE}

Similar to SCENT and MCE, SPIDE \cite{xu2022SPIDE} also defines an entropy for each cell using both gene expression levels and a PPI network. However, SPIDE uses a more complicated procedure to process gene expression levels. Since the notations in SPIDE's publication are erroneous, the following description of SPIDE is based on our understanding of both the publication and code implementation. 
Specifically, SPIDE first performs imputation to reduce data sparsity. In the code implementation, for each cell $j$, SPIDE finds its $k$-nearest neighbors ($k =10$ for datasets with $1{,}000$ cells or fewer; $k=25$ for datasets with more than $1{,}000$ cells) as those cells with the highest Pearson correlations with cell $j$ based on the $\log_2 (\text{count} + 1.1)$ transformed expression levels (the count of gene $i$ in cell $j$ is $X_{ij}$). However, as illustrated in Figure 1 of SPIDE's publication, SPIDE finds the $k$-nearest neighbors based on the cell-cell correlations calculated using log-transformed and quantile-normalized data, which is inconsistent with the code implementation.   
For each gene, if its raw count in cell $j$ is zero, then SPIDE computes two averages of the gene's raw count---one across cell $j$'s $k$-nearest neighbors and the other across all cells. If the first average is no less than the second (in the publication; in the code, "no less than" becomes "greater than"), SPIDE replaces the gene's zero expression level in cell $j$ with the first average. Then, SPIDE further processes the imputed data by quantile normalization and log transformation.

%Specifically, SPIDE performs quantile normalization on all cells followed by imputation to reduce data sparsity. For each cell $j$, SPIDE finds its $k$-nearest neighbors ($k =10$ for datasets with $1{,}000$ cells or fewer; $k=25$ for datasets with more than $1{,}000$ cells) as those cells with the highest Pearson correlations with cell $j$ based on the normalized gene expression levels. For each gene, if its normalized expression in cell $j$ is zero, then SPIDE computes two averages of the gene's normalized expression levels---one across cell $j$'s $k$-nearest neighbors and the other across all cells. If the first average is no less than the second, SPIDE replaces the gene's zero expression level in cell $j$ by the first average. 

With the imputed and log normalized gene expression levels and the PPI network, SPIDE defines the \textit{entropy} for cell $j$ as 
\[E_j = - \sum_{i=1}^p \Pi_{ij} \sum_{i^\prime \in N(i)} P_{ii^\prime j} \log P_{ii^\prime j}\,,\]
which is exactly the same formula as SCENT's signaling entropy. 

Despite the same entropy formula, there are three differences between SPIDE and SCENT. First, the two methods use different definitions of $Z_{ij}$, the expression level of gene $i$ in cell $j$, which is used to define $P_{ii^\prime j}$, the transition probability from gene $i$ to gene $i^\prime$ in cell $j$. Recall that SCENT defines $Z_{ij}$ as a log normalized expression level, while SPIDE defines $Z_{ij}$ by imputation followed by quantile normalization and log transformation. Second, SPIDE changes the definition of $W_{ii^\prime j}$ in $P_{ii^\prime j} = \frac{W_{ii^\prime j}}{\sum_{i'' \in N(i)}W_{ii''j}}$. Recall that in SCENT, $W_{ii^\prime j} = Z_{ij}Z_{i^\prime j}$, so  $P_{ii^\prime j} = \frac{Z_{i^\prime j}}{\sum_{i'' \in N(i)}Z_{i''j}}$. In contrast, SPIDE defines $W_{ii^\prime j} = \lvert R_{ii^\prime j} \lvert Z_{ij}Z_{i^\prime j}$, where $R_{ii^\prime j}$ the Pearson correlation between genes $i$ and $i^\prime$ based on the imputed, normalized, and log-transformed expression levels (in publication; in the code, $R_{ii^\prime j}$ is calculated using the  $\log_2(\text{count} + 1.1)$ transformed expression levels) in cell $j$'s $k$-nearest neighbors. Hence, $P_{ii^\prime j}$ in SPIDE becomes 
\[P_{ii^\prime j} = \frac{\lvert R_{ii^\prime j}\lvert Z_{i^\prime j}}{\sum_{i'' \in N(i)} \lvert R_{ii''j} \lvert Z_{i''j}}\,.\] 
Third, SPIDE changes the definition of $A$ in $\Pi_{ij} = \frac{Z_{ij}  (A \cdot Z_{\cdot j})_i}{Z_{\cdot j}^\intercal  \cdot  (A  \cdot  Z_{\cdot j})}$, the steady-state probability of gene $i$ in cell $j$. Recall that in SCENT, $A$ is the binary gene-gene adjacency matrix from the PPI network. In contrast, SPIDE changes $A$ to $A_j$, which becomes specific to cell $j$, and defines the $(i, i')$-th element of $A_j$ as
\[
A_{ii^\prime j} = \lvert R_{ii^\prime j}\lvert A_{ii^\prime}\,.
\]

Finally, SPIDE estimates the \textit{potency} for cell $j$ as \[\mathrm{\it{NE}}_j = E_j / C\,,\] where $C$ is exactly the same $C$ as in SCENT, i.e., $C = \log \lambda$, and $\lambda$ is the largest eigenvalue for the binarized adjacency matrix $A$. However, as SPIDE makes three changes in the definition of $E_j$, the same $C$ is no longer a valid normalization constant.

%In short, the major differences between SPIDE and SCENT are that: (1) SPIDE has an extra step for zero-imputation, (2) when calculating the transition probability, SPIDE also considered gene-gene correlation in addition to gene expression levels.

As SPIDE is conceptually similar to SCENT, they share the same intuition.
%The intuition behind SPIDE is that a cell is more stem-like if the identified influential genes (highly expressed genes that are densely connected in the PPI network and exhibit high correlation in expression levels) have more uniform weighted neighboring gene expression.  

The parameters in SPIDE include the offset value of $1.1$ used in the log transformation $Z_{ij} = \log_2(S_{ij} + \text{offset})$ and the number of nearest neighbors ($k$) of each cell. In total, we count SPIDE to have two parameters. \re{The default value choices of these two parameters were not justified by empirical evidence. In the \texttt{SPIDE} Python package (as of the update on June 22, 2022), the number of nearest neighbors is tunable by users, but the offset value for the log transformation is not.}

\section*{Category 3: correlation-based methods}

As the only three correlation-based methods, mRNAsi \cite{malta2018machine}, CCAT \cite{teschendorff2021ccat}, \re{and FitDevo \cite{zhang2022fitdevo}} estimate the \textit{potency} for each cell as the correlation between the cell's gene expression levels (a vector whose length is the number of genes) and a vector of genes' importance scores (pre-defined and not specific to any cells). 

\subsection*{mRNAsi}

As a method that applies to both bulk RNA-seq and scRNA-seq data, mRNAsi \cite{malta2018machine} trains a one-class logistic regression (OCLR) model, where genes are features (with gene expression levels as feature values), cells (or bulk samples) are observations (also known as instances), and the only class in training data contains bulk stem-cell samples only. From the trained model, mRNAsi extracts genes' coefficients as genes' importance scores. Finally, mRNAsi defines the \textit{potency} for each cell (or sample) in a new dataset as the Spearman correlation between the same genes' expression levels in the cell (or sample) and the importance scores. mRNAsi chooses the Spearman correlation instead of the Pearson correlation for better robustness to batch effects that differ between the training data and the new data.

The intuition behind mRNAsi is that a cell (or sample) has a higher potency if the rank of genes' expression levels in it is consistent with the genes' important scores obtained from the pre-trained OCLR model. 

There are no apparent parameters in mRNAsi.

\subsection*{CCAT}

Unlike mRNAsi, CCAT \cite{teschendorff2021ccat} uses the Pearson correlation instead of the Spearman correlation. Developed by the same group of authors, CCAT is a fast proxy of SCENT that remedies SCENT's scalability issue.  Instead of calculating cell-specific gene steady-state probabilities and gene-gene transition probabilities as SCENT does, CCAT estimates the \textit{potency} for each cell using the Pearson correlation between the log normalized gene expression levels\footnote{In CCAT, the log normalized expression of gene $i$ in cell $j$ is defined as $Z_{ij} = \log_2(S_{ij}+1)$, where $S_{ij}$ is $X_{ij}$ (the count of gene $i$ in cell $j$) multiplied by cell $j$'s specific scaling factor calculated externally. Note that the definition of $Z_{ij}$ is similar in CCAT and SCENT, except that the constant inside $\log_2$ is $1$ in CCAT and $1.1$ in SCENT.} in the cell and the genes' degrees in the inputted PPI network. However, the authors proved that CCAT approximates SCENT under an unrealistic assumption that every gene's neighboring genes have a constant expression level equal to the average expression level of all genes in each cell.

The intuition behind CCAT is that a cell has a higher potency if its gene expression levels has a stronger correlation with the genes' degrees in the PPI network; that is, the cell should have hub genes more highly expressed and non-hub genes more lowly expressed. However, the performance of CCAT is highly dependent on the inputted PPI network. If the PPI network is not specific to stem cells, then the hub genes may not be informative of potency.

There are no apparent parameters in CCAT.
\subsection*{FitDevo}
\re{Similar to CCAT, FitDevo \cite{zhang2022fitdevo} defines a cell's potency as the Pearson correlation between genes' importance scores, which are dataset-specific unlike in CCAT, and the expression levels of the same genes in the cell. FitDevo defines each gene's importance score through a complex procedure, calculating it as the sum of two binary scores: \textit{binarized gene weight (BGW)} and \textit{predicted binarized gene weight (pBGW)}. Note that the BGWs are pre-computed based on $17$ existing datasets and irrelevant to user-inputted scRNA-seq data, while the pBGWs are calculated based on the BGWs and user-inputted scRNA-seq data. The calculation of BGWs and pBGWs is detailed below.}

\begin{enumerate}
    \item \re{To calculate the BGWs, one per gene, FitDevo uses the $17$ datasets from CytoTRACE's publication \cite{gulati2020cytotrace}, where cells have discrete time-point labels. FitDevo calculates BGWs for $p=14{,}717$ genes, which are expressed in at least $9$ out of the $17$ datasets. These BGWs are stored and made available in FitDevo. The detailed calculation is described in the following sub-steps.}
    \begin{enumerate}
        \item[1.1.] \re{For each dataset $k=1,\ldots,17$ with $n^{(k)}$ cells, we use  $X_{ij}^{(k)}$ to indicate the expression level\footnote{The FitDevo publication \cite{zhang2022fitdevo} does not specify the unit of the expression level, so it is unclear whether the expression level is count, normalized value, or log normalized value.} of gene $i=1,\ldots,p$ in dataset $k$'s cell $j=1,\ldots,n^{(k)}$. Moreover, we use $T_j^{(k)}$ to denote the reverted time-point label for cell $j=1,\ldots,n^{(k)}$ in dataset $k$. The time points are reverted so that a potency-related gene, whose expression decreases from early to late time points, can have a large, positive correlation with the reverted time points. Then for each gene $i=1,\ldots,p$ and dataset $k=1,\ldots,17$, FitDevo computes $C_{ik}$, the Pearson correlation between gene $i$'s expression levels in dataset $k$'s $n^{(k)}$ cells $(X_{i1}^{(k)}, X_{i2}^{(k)}, \ldots, X_{in^{(k)}}^{(k)})\tran$ and the $n^{(k)}$ cells' reverted time-point labels $(T_1^{(k)}, T_2^{(k)}, \ldots, T_{n^{(k)}}^{(k)})\tran$.}
        \item[1.2.] \re{For each gene $i$, FitDevo standardizes each $C_{ik}$ as $C_{ik}^\prime = C_{ik} / \sqrt{\sum_{k^\prime=1}^{17} \frac{(C_{ik^\prime})^2}{17-1}}$.}
        \item[1.3.] \re{FitDevo defines the gene weight (GW) for gene $i$ by averaging the $17$ standardized Pearson correlations as $\mathrm{\it{GW}}_i = \sum_{k = 1}^{17} C^\prime_{ik} / 17$.}
        \item [1.4.] \re{For each gene $i$, FitDevo binarizes the GW by the sign to obtain the BGW: $\mathrm{\it{BGW}}_i = 1$ if $\mathrm{\it{GW}}_i > 0$; $\mathrm{\it{BGW}}_i = 0$ otherwise.}
    \end{enumerate}
    % \item \re{Since these $17$ datasets have known timepoint labels, the authors check the performance of GW in terms of cell potency estimation, which is defined as the Pearson correlation between GW and gene expression. Then, they also utilize binarized gene weight (BGW; $\text{BGW}_i = 1$ if $\text{GW}_i > 0$; $\text{BGW}_i = 0$ otherwise) to estimate cell potency and finds out its average performance across the $17$ training datasets is better then GW's. Thus, FitDevo uses the $14{,}717$ genes' BGW as the prior gene importance scores.}
    \item \re{To calculate the pBGWs, one per gene, FitDevo uses the pre-computed BGWs for the $p$ genes and the user-inputted scRNA-seq data with $n$ cells. The detailed calculation is described in the following sub-steps.}
    \begin{enumerate}
        \item [2.1.] \re{By default, FitDevo follows the Seurat V4 \cite{seuratv4} pipeline (now also compatible with the Seurat V5 \cite{seuratV5} pipeline) to obtain the top $L=50$ principal components (PC) from $Z$, i.e., the user-inputted scRNA-seq data after normalization, log transformation, highly variable gene selection, and scaling in Seurat. Specifically, $Z_{ij} = \log \left((X_{ij} / \sum_{i^\prime = 1}^p X_{i^\prime j}) \times 10{,}000 + 1\right)$, the $(i,j)$-th entry of $Z$, denotes the log normalized expression level of gene $i=1,\ldots,p$ in cell $j=1,\ldots,n$ of the user-inputted scRNA-seq data. We use $M_{lj}$ to note the $l$-th PC score of cell $j=1,\ldots,n$ in the user-inputted scRNA-seq data, $l = 1,\ldots,L$.}
        \item [2.2.] \re{
        FitDevo calculates $O_{il}$ as the Pearson correlation between gene $i$'s log normalized expression levels $(Z_{i1}, Z_{i2}, \ldots, Z_{in})\tran$ and the $l$-th PC $(M_{l1}, M_{l2}, \ldots, M_{ln})$, $i=1,\ldots,p$; $l=1,\ldots,L$.}
        \item [2.3.] \re{FitDevo fits a generalized linear model using the $p$ genes' BGWs as the response and the genes' $L$ correlations with the top $L$ PCs as the covariates: 
\[
    \begin{cases}
      BGW_i &\sim \mathrm{Binomial}(\mu_i)\,;\\
      \mathrm{logit}(\mu_i) &= \alpha + \sum_{l=1}^L \beta_l O_{il}\,.
    \end{cases}
\]}
       \item [2.4.] \re{The parameters $\alpha, \beta_1, \dots,  \beta_L$ are estimated as  $\hat{\alpha}, \hat{\beta}_1, \dots, \hat{\beta}_L$. Then, the predicted $\mathrm{logit}(\mu_i)$ for gene $i$ is $\mathrm{logit}(\hat{\mu}_i) = \hat{\alpha} + \sum_{l=1}^L \hat{\beta}_l O_{il}\,$. FitDevo defines the predicted binarized gene weight (pBGW) as
\[
    pBGW_i = 
    \begin{cases}
    1, &\text{if } \mathrm{logit}(\hat{\mu}_i) > 0\,; \\
    0, &\text{otherwise}\,.
    \end{cases}
\]}
    \end{enumerate}
\end{enumerate}

\re{With the $p$ genes' BGWs and pBGWs, FitDevo defines the sample-specific gene weight (SSGW) for gene $i=1,\ldots,p$ by adding up gene $i$'s BGW and pBGW: $SSGW_i = BGW_i + pBGW_i$.} 

\re{Finally, FitDevo defines the \textit{potency} for a particular cell $j$ in the user-inputted scRNA-seq data as the Pearson correlation between all $p$ genes' log normalized expression levels in this cell $(Z_{1j}, Z_{2j}, \ldots, Z_{pj})\tran$ and their corresponding SSGWs $(SSGW_{1}, SSGW_{2}, \ldots, SSGW_{p})\tran$.}

\re{The only parameter in FitDevo is the number of PCs used ($K$) when calculating pBGWs. The default value of $K$ is set to the default value in Seurat's \texttt{RunPCA()} function. This default value choice of $K$ was not justified by empirical evidence in FitDevo's publication. It is a tunable parameter in the \texttt{FitDevo} R script (v1.0.1 and all subsequent versions).}

\re{Additionally, although FitDevo depends on the Seurat pipeline \cite{seuratv4}, it includes a ``no normalization" option, allowing users to input already-normalized data that do not require further normalization.}

\section*{Discussion}
In this review, we summarized \re{$14$} cell potency estimation methods into three categories, including \re{average-based} methods, entropy-based methods, and correlation-based methods. For each \re{of the $14$} method, we provided a high-level intuition and used a unified set of mathematical notations to detail the procedure. We also summarized the \re{$14$} methods' usage complexities, including the number of parameters, the number of required inputs, and the existence of \re{inconsistencies} between the method description in publications and the method implementation in software packages in Figure~\ref{fig:radar}. \re{Last, we summarized the differences among these $14$ methods regarding their inputs and outputs in Table~\ref{tab:summary}}.

\re{By closely examining the 14 methods, we identified some biological assumptions that warrant further discussion. First, as many of the $14$ methods reviewed here require users to input a PPI network,} an issue worth noting is the choice of PPI network. We found CCAT's robustness statement somewhat counter-intuitive: ``\textit{the robust association of CCAT with differentiation potency is due to the subtle positive correlation between transcriptome and connectome, which, as shown, is itself robust to the choice of PPI network}'' \cite{teschendorff2021ccat}. \re{If different PPI networks lead to the same potency estimates, it raises questions about the specific contribution of any particular PPI network.} \re{Second,} another issue worth discussing is the fundamental assumption of CytoTRACE that the gene count (number of genes expressed) reflects a cell's potency. Given that a cell's gene count is often highly correlated with its library size due to the sampling process, where typically only about 10-20\% of transcripts can be captured by scRNA-seq, we can deduce that a cell's library size may indicate its potency. \re{Conversely, the total amount of transcripts in the cells is not necessarily relevant to high potency. For instance, B cell activation results in a nearly 10-fold increase in total RNA \cite{libsize1}. Additionally, there is significant variability in total mRNA counts at both the organ and cell type levels \cite{libsize2}, which does not imply a difference in potency. Thus, the relationship between cell library size and cell potency remains unclear.} \re{Third, and most importantly, we found that certain methods might have contradictory intuitions. Two average-based methods, CytoTRACE and NCG, define a cell's potency based on the expression levels of a set of ``important" genes, with each method using its own criteria to determine importance, while underweighting the expression levels of other genes. Similarly, correlation-based methods assign high potency to a cell if the cell's highly expressed genes have similarly high importance scores, which are typically pre-defined. In contrast, some entropy-based methods, such as stemID, dpath, scEnergy, and cmEntropy, operate on the intuition that a cell has higher potency when its genes have similar expression levels, without explicit specification of gene importance.}

Moreover, we would like to point out some methodological \re{issues}. The first \re{issue} is with SLICE's application of the $k$-means clustering algorithm on a gene-gene dissimilarity matrix (a symmetric matrix) to find gene groups. \re{In the standard use of k-means clustering algorithm, the Euclidean distance should be defined based on an instance-by-feature matrix, where instances correspond to genes in this context, and features refer to characteristics of instances (e.g., genes’ GO terms), instead of an instance-by-instance dissimilarity matrix.} In SLICE's use of the $k$-means clustering algorithm, however, it is difficult to justify why the Euclidean distance defined for two genes based on a gene-gene dissimilarity matrix is reasonable. The second \re{issue} is with dpath's use of the Poisson distribution to model the log-transformed TPM data, which are not counts and thus cannot follow the Poisson distribution. Additionally, these methods employ various approaches to normalize gene expression levels, and it is unclear which normalization approach is more appropriate for cell potency estimation. 

Observing the lack of consensus \re{in method assumptions and approaches}, \re{along with new advancements in cell potency estimation from more perspectives (e.g., CytoTRACE2 using deep learning \cite{cytotrace2} and stemFinder using cell-cycle gene expression \cite{cellCyclePotency}),} we regard the task of estimating cell potency from scRNA-seq data as an open challenge for two reasons. 
\re{First, comprehensive method benchmarking with biological interpretation is challenging. The $14$ methods' internal benchmarking was limited because the datasets used (Supplementary Table S2) only contained time points or cell types with a known differentiation order, but did not provide single-cell-level cell potency ground truths.} Given that no available scRNA-seq data contains single-cell-level cell potency ground truths, thorough benchmarking of methods remains challenging. One possible direction is to use lineage tracing data \cite{lineageTracing, lineageTracing2}, which captures the differentiation lineages of cells. Another complementary direction is to use real-data-based synthetic null data to quantify the possible biases in cell potency estimation \cite{clusterDE}. \re{Second, the $14$ methods' internal benchmarking results showed a large discrepancy in the methods' performance on different datasets, as shown in Figure 2C of the CytoTRACE publication \cite{gulati2020cytotrace} and Figure 3C, 3D of the FitDevo publication \cite{zhang2022fitdevo} (see Supplementary Figures~\ref{fig:cytotrace_2c} and~\ref{fig:fitdevo_3}). However, it remains to investigate what data characteristics make certain methods suitable. Given the conceptual differences among the methods, such as whether gene importance is considered or not, it is imperative to understand the connections between data characteristics and method assumptions.} 

\re{In addition to furthering our understanding of cell potency estimation, a broader open question is regarding the distinct advantages of computational methods that use different perspectives (pseudotime inference, RNA velocity inference, and cell potency estimation) to infer cellular dynamics from static scRNA-seq data. In particular, there is a lack of consensus on how the three types of computational methods should be used jointly, or whether one or more types are more suitable for certain biological contexts. In some cases, the outputs from the three types of methods are treated as equivalent. For example, CellRank2 \cite{cellrank2} accepts outputs of all three types of methods for constructing a cell-cell transition matrix, which it uses for downstream inference of cell states and cell fate probabilities. Additionally, Qin et al.~utilize the sum of the estimated potency from CCAT and the magnitude of RNA velocity from scVelo to represent the pluripotency of each cell. In other cases, the outputs of different types of methods are used sequentially. For example, the estimated cell potency serves as a proceeding step to facilitate cell trajectory inference in two ways: (1) some cell potency estimation methods (stemID, SLICE, dpath, scEnergy, and SCENT) construct cell trajectories based on their estimated cell potency; (2) other methods use cell potency to identify the root cell in already-constructed trajectories. These different usages may hinder the reproducibility of scRNA-seq data analysis results. Therefore, a comprehensive benchmarking and a deeper understanding of these three types of methods are necessary to improve the trustworthiness of discoveries made from scRNA-seq data.}

\re{Looking ahead, we believe integrating more well-established potency-related biological knowledge can enhance computational methods' ability to quantify cell potency and uncover biological insights. First, leveraging well-characterized potency-related gene regulatory networks (GRNs) can possibly provide a more accurate estimation of cell potency, as a cell's pluripotency is known to be influenced by regulatory activities among genes \cite{GRNreview, GRNwithHSC, GRNwithESC}. Second, incorporating the cell-cell interaction information may improve the accuracy of cell potency quantification, as stem cells reside in fixed anatomical areas called niches, and activities like cell-cell interactions inside niches influence the proliferation and differentiation of stem cells \cite{niche, HSCmaintain, ottone2014direct}. Last, utilizing single-cell ATAC-seq data alongside scRNA-seq data may enhance computational methods' performance in quantifying cell potency, as chromatin accessibility and transcription factors' activities change during the cell differentiation process \cite{otherOmics, chromatin1, chromatin2}.}

\re{In conclusion, we envision that experimental and computational approaches can work cooperatively to enhance the ability to quantify cell potency and identify more potency marker genes across various tissues and conditions. As a performance check, a robust cell potency estimation method should be able to recognize known potency marker genes. Such methods can then be used to discover novel marker genes, which biologists can experimentally validate (e.g., by knocking out these potential marker genes \cite{crisprMarker1, crisprMarker2}). If these novel genes are proven to be potency-related, they can serve as new features to help quantify cell potency. Additionally, with increased knowledge of potency-related marker genes, biologists can further investigate the key signaling pathways and regulatory activities among these genes. This cooperation between computational algorithms and experimental validation can lead to more reliable and insightful findings regarding cellular differentiation.}

\re{To facilitate the single-cell community's future benchmark studies, we have curated Supplementary Table S2, which includes the datasets used in these cell potency estimation methods, although these datasets do not provide single-cell-level cell potency ground truths.}

\subsection*{Data Availability}
No data are associated with this article.

\subsection*{Author contributions}
Q.W., Z.Z., D.S., and J.J.L. conceived the project. Q.W., Z.Z., and J.J.L. reviewed the cell potency estimation methods. Q.W., Z.Z., Q.L., D.S., and J.J.L. discussed, drew, and revised the figures. Q.W. summarized the datasets mentioned in the original methods' publications. Q.W. and J.J.L. wrote the manuscript. Z.Z., Q.L., D.S., and J.J.L. revised the manuscript. J.J.L. supervised the project. All authors participated in discussions and approved the final manuscript.

\subsection*{Competing interests}
The authors declare no competing interests.

\subsection*{Grant information}
This work is supported by NSF DGE-2034835 to Q.W. and NIH/NIGMS R35GM140888, NSF DBI-1846216 and DMS-2113754, Johnson \& Johnson WiSTEM2D Award, Sloan Research Fellowship, UCLA David Geffen School of Medicine W.M. Keck Foundation Junior Faculty Award, and the Chan-Zuckerberg Initiative Single-Cell Biology Data Insights to J.J.L.

\subsection*{Acknowledgements}
The authors appreciate the comments and feedback from the Junction of Statistics and Biology members at UCLA (http://jsb.ucla.edu). This material is based upon work supported by the National Science Foundation Graduate Research Fellowship Program under Grant No.DGE-2034835. Any opinions, findings, and conclusions or recommendations expressed in this material are those of the author(s) and do not necessarily reflect
the views of the National Science Foundation.

\newpage

{\small\bibliographystyle{unsrtnat}

}

\clearpage
\section*{Figures}

\begin{figure}[htbp]
    \centering
    \includegraphics[width = \textwidth]{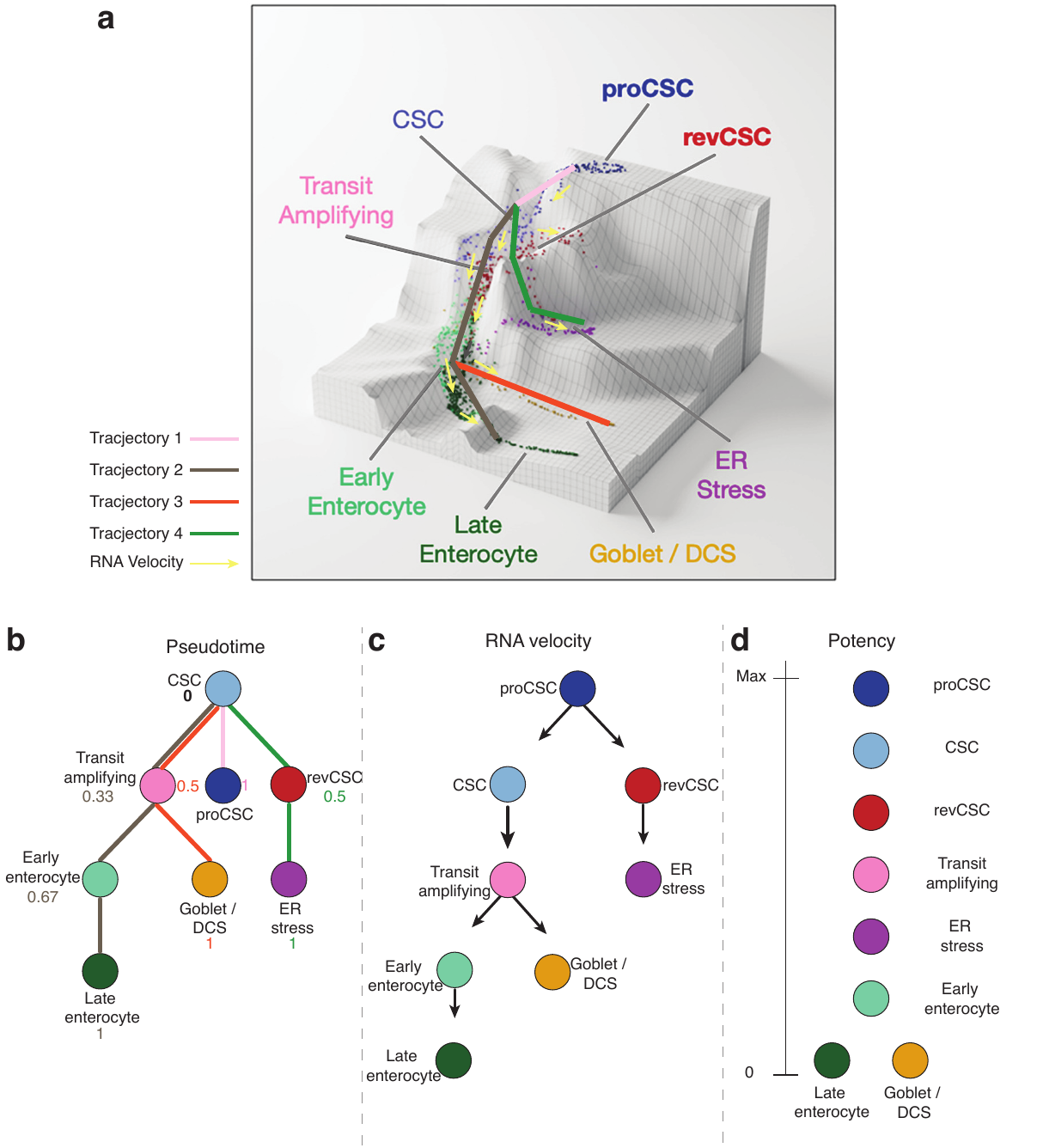}
\end{figure}

\afterpage{
\clearpage
\begin{figure}
    \caption{\re{Different information provided by pseudotime, RNA velocity, and potency. \textbf{a}) Combined illustration of pseudotime, RNA velocity, and cell potency. This figure is recreated based on a part of Figure 7A from Qin et al.'s open access publication \cite{3dFigure}, and this recreation is permitted under the open access terms. The dots with different colors represent cells from different cell types, which include colonic stem cells (CSC), oncogene-driven LRIG1$^+$ hyper-proliferative colonic stem cells (proCSC), fibroblast-induced Clusterin (CLU)$^+$ revival colonic stem cells (revCSC), transit amplifying cells (transit amplifying), cells under endoplasmic reticulum stress (ER stress), early enterocytes, late enterocytes, and goblet and deep crypt secretory cells (goblet/DCS). The brown, pink, green, and orange lines represent four differentiation trajectories, which can be inferred using pseudotime inference methods with CSC specified as the root cell type. The yellow arrows represent RNA velocity vectors, which can be inferred using RNA velocity inference methods. The altitude of this three-dimensional plot reflects cell potency, which can be inferred using cell potency estimation methods. \textbf{b}) Illustration of pseudotime inference results, containing lineages (represented by lines of different colors) and pseudotime values within each lineage (represented by numerical values from 0 to 1, with 0 indicating the root cell). Colored circles represent cell types. \textbf{c}) Illustration of RNA velocity inference results, with RNA velocities represented by arrows indicating direction and varying lengths indicating velocity magnitudes. \textbf{d}) Illustration for cell potency estimation results, where potency values are numeric and can be used to rank the cells or cell types. Note that there is a discrepancy between the pseudotime results in \textbf{b} and the RNA velocity and cell potency results in \textbf{c} and \textbf{d}. In \textbf{b}, CSC is specified as the root cell type based on domain knowledge, while in \textbf{c} and \textbf{d}, proCSC is an ancestor of CSC in terms of RNA velocity and cell potency. According to Qin et al.'s publication \cite{3dFigure}, proCSC is oncogene-driven and emerges subsequent to CSC. Therefore, the pseudotime results based on domain knowledge are considered more reasonable, suggesting that RNA velocity and cell potency results that do not incorporate domain knowledge might be problematic in some cases.}}
    \label{fig:3d_compare}
\end{figure}
}
\clearpage

% \begin{figure}[h]
% \centering
% \includegraphics[width=0.9\textwidth]{Waddingtons-landscape-model-From-Waddingt-on-1957-29.png}
% \caption{\label{fig:waddington}Waddington's landscape \cite{waddington1957}. This ball symbolizes a cell undergoing a differentiation process, while the valleys symbolize possible differentiation trajectories. As the ball descends the hill, it selects a specific pathway and eventually rests at the bottom of the hill, representing a differentiated cell that reaches its stable state.}

% \end{figure}

\begin{figure}
    \centering
    \includegraphics[width=\textwidth]{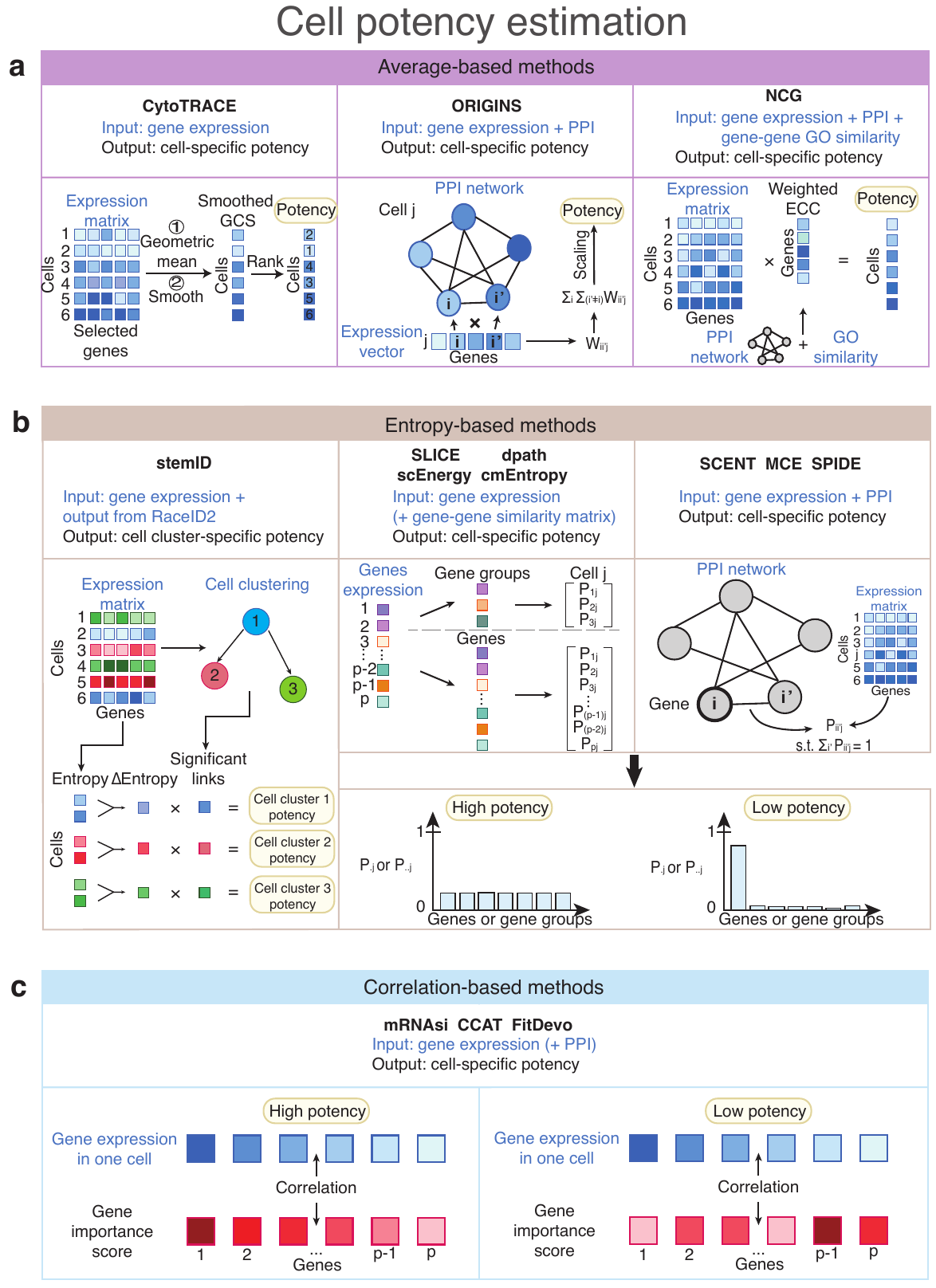}
\end{figure}

\afterpage{
\begin{figure}
    \caption{A conceptual categorization of the \re{$14$} methods into three categories: \re{average-based} methods, entropy-based methods, and correlation-based methods. All methods except stemID calculate the potency at the single-cell resolution, while stemID offers the cell-cluster resolution only. \textbf{a)} \re{Average-based} methods. The CytoTRACE method (left) computes a Gene Count Signature (GCS) for each cell using the selected genes only. Then CytoTRACE defines cells' potency as the ranks of the cells' smoothed GCS values. \re{The ORIGINS method (middle) assigns every edge in the PPI network a cell-specific weight, calculated as the product of the cell-specific expression levels of the two genes connected by the edge. ORIGINS defines a cell's potency as a scaled sum of these cell-specific edge weights.} The NCG method (\re{right}) defines a weight for each gene by aggregating the gene's similarities to other genes. These gene-gene similarities are based on the genes' edge weights in a PPI network and their GO terms. NCG then uses the defined gene weights and a gene-by-cell log-normalized expression matrix to estimate each cell's potency. \textbf{b)} Entropy-based methods. These \re{eight} methods use entropy to define the potency and can be divided into three subcategories. In the first subcategory, stemID defines the potency of each cell cluster, not each cell, as the product of the cell cluster's relative entropy ($\Delta \text{Entropy}$) and links (similarities) to other cell clusters. In the second subcategory, SLICE, dpath, scEnergy, \re{and cmEntropy} first calculate a proportion vector $P_{\cdot j}$ for each cell $j$, where $P_{ij}$ is the proportion (i.e., the relative expression level) of gene group $i$ (SLICE and dpath) or gene $i$ (scEnergy \re{and cmEntropy}) in cell $j$, and then define cell $j$'s potency as the entropy of $P_{\cdot j}$. Among the \re{four} methods in the second sub-category, SLICE requires users to input a gene-gene similarity matrix (the Kappa statistics for all gene pairs) while the other \re{three} methods do not. In the third subcategory, SCENT, MCE, and SPIDE define cell $j$'s potency as the entropy of a transition probability matrix (or extended vector) $P_{\cdot \cdot j}$, where $P_{ii^\prime j}$ is the transition probability (i.e., the normalized edge weight) from gene $i$ to gene $i^\prime$ in cell $j$. \textbf{c)} Correlation-based methods. The \re{mRNAsi, CCAT, and FitDevo} methods define a cell's potency as the correlation between the cell's gene expression vector and a vector of pre-defined, non-cell-specific gene important scores. In mRNAsi, genes' importance scores are the genes' coefficients in a pre-trained one-class logistic regression model for predicting the probability that a given cell is a stem cell. In CCAT, genes' importance scores are the genes' degrees in a PPI network. \re{In FitDevo, a gene' importance score is the sum of a pre-trained gene weight and a sample-specific gene weight.} CCAT requires users to input a PPI network while mRNAsi \re{and FitDevo do} not.}
    \label{fig:cellpotency}
\end{figure}
}

\clearpage
\begin{figure*}[h]
    \centering
    \includegraphics[width=\textwidth]{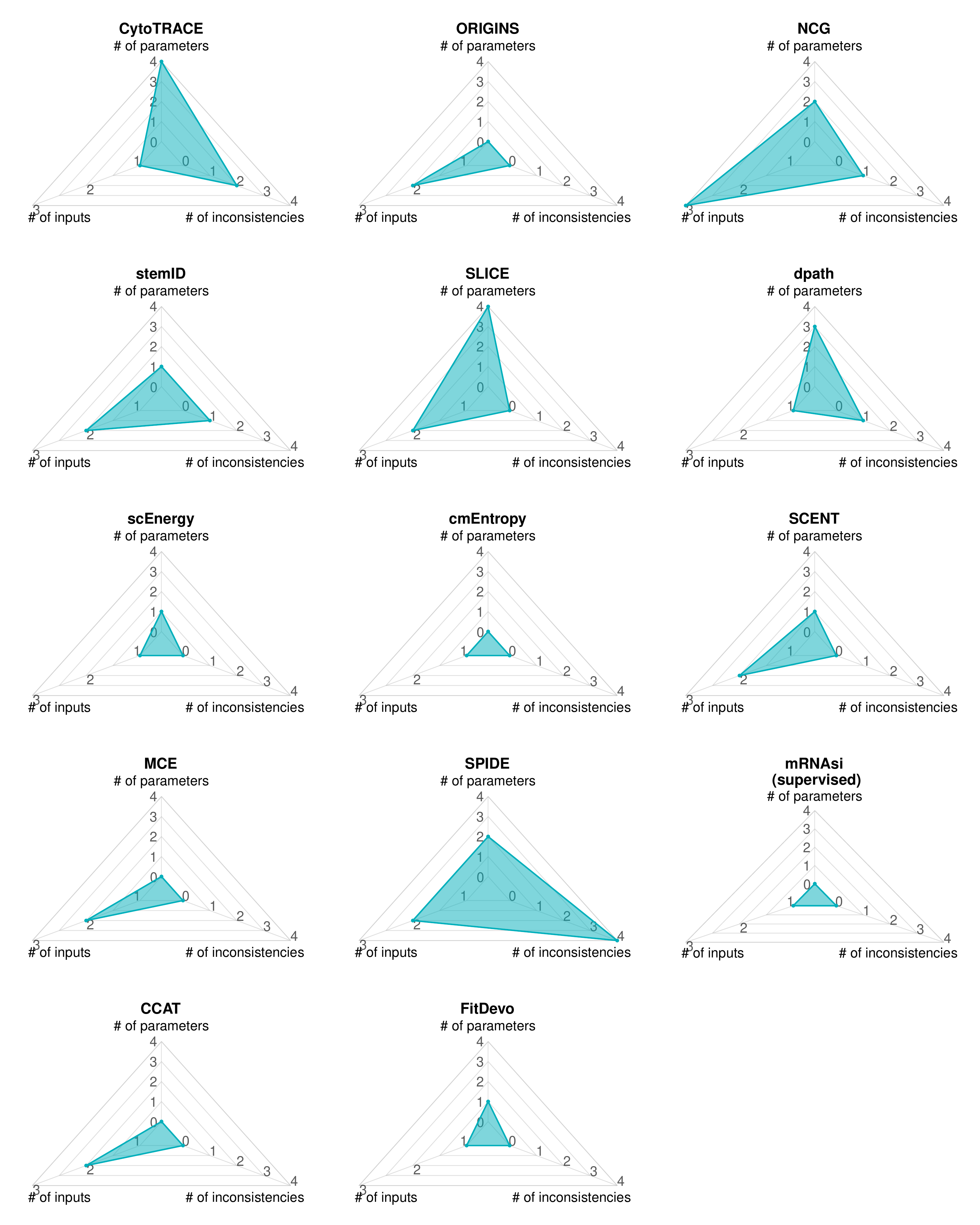}
    \caption{A radar plot of the \re{14} cell potency estimation methods, illustrating their number of inputs, number of parameters, and number of \re{inconsistencies} between publication and code implementations, reflecting the simplicity (\re{fewer inputs and parameters}) and \re{consistency (fewer inconsistencies)} of the methods.}
    \label{fig:radar}
\end{figure*}
\clearpage

\section*{Tables}
\clearpage

\begin{table}[]
\centering
\caption{Comparisons of the inputs and outputs for pseudotime inference, RNA velocity inference, and cell potency estimation methods.}
\label{input_output}
\begin{tabular}[t]{lll}
\toprule
Category       & Input                                                                                                                                                & Output                                                                                                                                                                                          \\ \midrule
Pseudotime inference  & \begin{tabular}[c]{@{}l@{}}Low-dimensional cell embeddings \\ of the scRNA-seq data; \\ cells' cluster or type labels; \\ the root cell\end{tabular}                          & Cell-by-trajectory  pseudotime matrix \\ \newline \\
RNA velocity inference & \begin{tabular}[c]{@{}l@{}}Spliced and  unspliced gene \\ expression matrices\end{tabular}                                                           & Cell-by-gene velocity matrix  \\  \newline \\
Potency estimation     & \begin{tabular}[c]{@{}l@{}}Gene-by-cell expression matrix; \\ (optional: PPI network; gene-gene \\ similarity matrix; cell cluster labels)\end{tabular} & Cell potency vector               \\ \bottomrule
\end{tabular}
\end{table}

\begin{sidewaystable}
\begin{center}
\caption{Summary of cell potency estimation methods}\label{tab:summary}
\begin{adjustbox}{width=\linewidth,center}
\renewcommand{\arraystretch}{2.5}
\begin{tabular}[t]{lllll}
\toprule
Method      & Potency measure                            & Input                                                                                      &       Cell potency resolution  &    Implementation (version or update date when accessed)          \\
\midrule
CytoTRACE \cite{gulati2020cytotrace} & \re{Average$^\ast$}        & Count matrix$^\dagger$                                                                                                                                                                               & Single-cell level & R package \href{https://cytotrace.stanford.edu/}{\texttt{CytoTRACE} (v 0.3.3)}                                                                   \\
\re{ORIGINS \cite{senra2022origins}} & \re{Average$^\ast$}   & \re{Count matrix$^\dagger$; PPI network} & \re{Single-cell level } & \href{https://github.com/danielasenraoka/ORIGINS/tree/main}{\re{R package \texttt{ORIGINS} (v 0.1.0)}}\\
NCG \cite{ni2021accurate}   & \re{Average$^\ast$} & Count matrix$^\dagger$; PPI network; gene-gene similarity matrix                                                 & Single-cell level  & \href{https://github.com/Xinzhe-Ni/NCG}{R script (September 1, 2021)}                                                                         \\

stemID \cite{grun2016stemID}    & Entropy                              & Count matrix$^\dagger$; Output from RaceID2                                                                                                                                                                           & Cell-cluster level     & \href{https://github.com/dgrun/StemID}{R script (March 3, 2017)}                                                                       \\
SLICE \cite{guo2017slice}    & Entropy                              & Normalized expression matrix$^\ddagger$; gene-gene similarity matrix                                                                                                                                                                      & Single-cell level & R package \href{https://github.com/xu-lab/SLICE}{\texttt{SLICE} (v 0.99.0)}                                                                        \\
dpath \cite{gong2017dpath}    & Entropy                              & TPM matrix                                                                                                                                                                                            & Single-cell  level & R package \href{https://www.nature.com/articles/ncomms14362#MOESM1580}{\texttt{dpath} (in Supplementary Software 1; v 2.0.1)}                                                                    \\
scEnergy \cite{jin2018scEnergy}  & Entropy                              & Normalized expression matrix$^\ddagger$                                                                                                                                                                                           & Single-cell level  & \href{https://github.com/sqjin/scEpath}{MATLAB script (February 16, 2021)}                                                                    \\

\re{cmEntropy \cite{kannan2021cmEntropy}}&\re{Entropy} &\re{Count matrix$^\dagger$} & \re{Single-cell level}& \href{https://github.com/skannan4/cm-entropy-score}{\re{R script (December 30, 2020)}}\\

SCENT \cite{teschendorff2017SCENT}    & Entropy                              & Log normalized expression matrix$^\S$; PPI network                                                                       & Single-cell level  & R package \href{https://github.com/aet21/SCENT}{\texttt{SCENT} (v 1.0.3)}                                                                      \\

MCE \cite{shi2020mce}      & Entropy                              & Count matrix$^\dagger$; PPI network                                                                  & Single-cell level  & \href{https://academic.oup.com/bib/article/21/1/248/5115275#supplementary-data}{MATLAB script (in Supplementary data; January, 2020)}                                                                    \\

SPIDE \cite{xu2022SPIDE}     & Entropy                              & Count matrix$^\dagger$; PPI network                                                               & Single-cell level  & Python package \href{https://github.com/CSUBioGroup/SPIDE}{\texttt{SPIDE} (June 22, 2022)}                                                                 
\\
mRNAsi \cite{malta2018machine}   & Correlation        & Count matrix$^\dagger$                                                                                                                            & Single-cell level  & \href{https://github.com/BioinformaticsFMRP/PanCanStem_Web}{R script (December 2, 2021)}                                                                               \\

CCAT \cite{teschendorff2021ccat}     & Correlation                  &  Log normalized expression matrix$^\S$; PPI network                                                                       & Single-cell level & R package \href{https://github.com/aet21/SCENT}{\texttt{SCENT} (v 1.0.3)}  \\

\re{FitDevo \cite{zhang2022fitdevo}}& \re{Correlation}& \re{Count matrix$^\dagger$}  & \re{Single-cell level}& \href{https://github.com/jumphone/fitdevo}{\re{R script (April 6, 2024)}}\\
\bottomrule
\end{tabular}
\end{adjustbox}
\end{center}
\footnotesize{$\ast$: Subject to weighting and smoothing.}
\newline
\footnotesize{$^\dagger$: A gene-by-cell count matrix processed from scRNA-seq reads, denoted by $X$.}
\newline
\footnotesize{$^\ddagger$: A gene-by-cell normalized expression matrix, denoted by $S$.}\\
\footnotesize{$^\S$: A gene-by-cell log normalized expression matrix, denoted by $Z$.}
\end{sidewaystable}
\clearpage

\begin{table}[]
\centering
\caption{Intuitions underlying the \re{$14$} cell potency estimation methods}\label{intuition}
\begin{tabular}{lll}
\toprule
Method    & Potency measure & Intuition (what cells have higher potency)\\
\midrule
CytoTRACE &\re{Average} & Cells with \re{more genes expressed} \\
\newline\\
\re{ORIGINS} & \re{Average} &  \begin{tabular}[l]{@{}l@{}}\re{Cells where a larger number of highly expressed genes} \\\re{are connected in a PPI network}\end{tabular}\\ 
\newline\\
NCG       &\re{Average} & \begin{tabular}[l]{@{}l@{}}Cells with higher expression levels of a specific subset \\
of genes, which are closely connected in a PPI \\network and share similar GO terms\end{tabular} \\
\newline\\

stemID    &Entropy & \begin{tabular}[l]{@{}l@{}}Cell clusters that have more neighboring clusters \\and more cells with \re{genes exhibiting similar expression levels}\end{tabular}   \\
\newline\\
SLICE     &Entropy & \begin{tabular}[l]{@{}l@{}}Cells \re{where gene groups have more similar proportions } \\\re{of expressed genes}\end{tabular}\\
\newline\\
dpath    &Entropy  & \begin{tabular}[l]{@{}l@{}}Cells where metagenes exhibit more \re{similar} expression levels\end{tabular}\\
\newline\\
scEnergy  &Entropy & \begin{tabular}[l]{@{}l@{}}Cells where genes exhibit more \re{similar} expression levels\end{tabular}\\
\newline\\
\re{cmEntropy}& \re{Entropy}& \re{Cells where genes exhibit more \re{similar} expression levels} \\
\newline\\
SCENT     &Entropy & \begin{tabular}[l]{@{}l@{}}Cells whose influential genes (highly expressed genes\\ that are densely connected with other highly expressed \\genes in a PPI network) have neighboring genes \\ \re{expressed at more similar levels} \end{tabular}\\
\newline\\
MCE      &Entropy  & \begin{tabular}[l]{@{}l@{}}Cells with more \re{similar} weighted gene-gene transition probabilities \\\end{tabular}\\
\newline \\
SPIDE    &Entropy  & \begin{tabular}[l]{@{}l@{}}Cells whose influential genes (highly expressed genes\\ that are highly correlated and densely connected with \\other highly expressed genes in a PPI network) 
have \\neighboring genes \re{expressed at more similar levels}
\\  \end{tabular}                     \\
\newline\\
mRNAsi   &Correlation & \begin{tabular}[l]{@{}l@{}}Cells that have a more consistent ranking between gene \\
expression levels and pre-trained gene importance scores\end{tabular}                                                              \\
\newline\\
CCAT     &Correlation & \begin{tabular}[l]{@{}l@{}}Cells that have a stronger positive correlation \\between genes' expression levels and their degrees\\in a PPI network\end{tabular}\\
\newline \\
\re{FitDevo} &\re{Correlation} & \re{\begin{tabular}[l]{@{}l@{}}Cells whose gene expression levels have a stronger \\ correlation with pre-defined gene importance scores\end{tabular}}\\
\bottomrule
\end{tabular}
\end{table}
\clearpage

\begin{table}[]
\centering
\caption{\re{Parameters of the $14$ methods.}}
\label{tab:total_parm}
\begin{tabular}{lll}
\toprule
Method    & Parameters                                                                                                                                                                                                                       & \begin{tabular}[c]{@{}l@{}} Total number of\\ parameters\end{tabular} \\
\midrule
CytoTRACE & \begin{tabular}[c]{@{}l@{}}Number of genes used to calculate GCS ($200$);\\ Number of genes with the largest dispersion indices ($1{,}000$);\\ Percentage  threshold when filtering genes ($5\%$);\\ $\alpha$ in the diffusion process ($0.9$);\end{tabular} & 4                          \\
\newline \\
ORIGINS   & None                                                                                                                                                                                                                             & 0                          \\
\newline \\
NCG       & \begin{tabular}[c]{@{}l@{}}Offset used in log transformation ($1.1$);\\ Cutoff value for $s_{ij}$ ($1$)\end{tabular}                                                                                                                           & 2                          \\
\newline \\

stemID    & Significant level for identifying significant links ($0.01$)                                                                                                                                                                              & 1                          \\
\newline \\
SLICE     & \begin{tabular}[c]{@{}l@{}}Gene expression levels threshold for filtering genes ($1$);\\ Size of the sub-sample $G_b$($p_s$);\\ Number of gene groups ($M$);\\ Number of sub-samples ($B$)\end{tabular}                                                      & 4                          \\
\newline \\
dpath     & \begin{tabular}[c]{@{}l@{}}Number of metagenes ($M$);\\ Mean parameter for the Poisson distribution for the dropout ($0.1$);\\ $\pi_{ij}$ in the first round of NMF for $Z_{ij} = 0$ ($0.1$)\end{tabular}                                                                  & 3                          \\
\newline \\
scEnergy  & Correlation threshold when constructing the gene network ($0.1$)                                                                                                                                                                         & 1                          \\
\newline \\
cmEntropy & None                                                                                                                                                                                                                             & 0                           \\
\newline \\
SCENT     & Offests in log transformation ($1.1$)                                                                                                                                                                                                  & 1                          \\
\newline \\
MCE       & None                                                                                                                                                                                                                             & 0                          \\
\newline \\
SPIDE     & \begin{tabular}[c]{@{}l@{}}Offest in log transformation ($1.1$);\\ k in finding a cell's k-nearest neighbors ($10$ or $25$)\end{tabular}                                                                                                                & 2                          \\
\newline \\
mRNAsi    & None                                                                                                                                                                                                                             & 0                          \\
\newline \\ 
CCAT      & None                                                                                                                                                                                                                             & 0                          \\
\newline\\
FitDevo   & Number of PCs ($K$)                                                                                                                                                                                                                             & 1  \\
\newline \\
\bottomrule 
\end{tabular}
\end{table}
\clearpage

\begin{table}[]
\centering
\caption{\re{Parameters with unjustified default values of the $14$ methods. Among these parameters, the tunable parameters are listed.}}
\begin{adjustbox}{width=\textwidth}
\begin{tabular}{lllll}
\toprule
           & \begin{tabular}[c]{@{}l@{}}Parameters with \\ unjustified default values\end{tabular}                                                                                                         & \begin{tabular}[c]{@{}l@{}}\# of parameters \\ with unjustified \\ default values\end{tabular} & \begin{tabular}[c]{@{}l@{}}Tunable \\ parameters\end{tabular}                                                                                                                     & \begin{tabular}[c]{@{}l@{}}\# of tunable \\ parameters\end{tabular} \\
\midrule
CytoTRACE & \begin{tabular}[c]{@{}l@{}}Number of dispersed \\ genes kept after filtering;\\ Percentage  threshold \\ when filtering genes\end{tabular}                                                  & 2                                                                                            & None                                                                                                                                                                              & 0                                                                   \\
\newline \\
ORIGINS    & None                                                                                                                                                                                        & 0                                                                                            & None                                                                                                                                                                              & 0                                                                   \\
\newline \\
NCG        & \begin{tabular}[c]{@{}l@{}}Offset used in log \\ transformation;\\ Cutoff value for $s_{ij}$\end{tabular}                                                                                   & 2                                                                                            & None                                                                                                                                                                              & 0                                                                   \\
\newline \\

stemID     & \begin{tabular}[c]{@{}l@{}}Significant level for \\ identifying significant \\ links\end{tabular}                                                                                           & 1                                                                                           & \begin{tabular}[c]{@{}l@{}}Significant level for \\ identifying significant links\end{tabular}                                                                                    & 1                                                                   \\
\newline \\
SLICE      & \begin{tabular}[c]{@{}l@{}}Thresholding gene \\ expression levels for \\ filtering genes;\\ Size of the sub-sample;\\ Number of gene groups;\\ Number of sub-samples\end{tabular}           & 4                                                                                            & \begin{tabular}[c]{@{}l@{}}Thresholding gene \\ expression levels for \\ filtering genes;\\ Size of the sub-sample;\\ Number of gene groups;\\ Number of sub-samples\end{tabular} & 4                                                                   \\
\newline \\
dpath      & \begin{tabular}[c]{@{}l@{}}Number of metagenes;\\ Mean parameter for the \\ Poisson distribution for\\ the dropout;\\ $\pi_{ij}$ in the first round \\ of NMF for $Z_{ij} = 0$\end{tabular} & 3                                                                                            & \begin{tabular}[c]{@{}l@{}}Number of metagenes;\\ Mean parameter for the \\ Poisson distribution for \\ the dropout;\end{tabular}                                                 & 2                                                                   \\
\newline \\
scEnergy   & \begin{tabular}[c]{@{}l@{}}Correlation threshold \\ when constructing the \\ gene network\end{tabular}                                                                                      & 1                                                                                           & \begin{tabular}[c]{@{}l@{}}Correlation threshold \\ when constructing the \\ gene network\end{tabular}                                                                            & 1                                                                   \\
\newline \\
cmEntropy  & None                                                                                                                                                                                        & 0                                                                                            & None                                                                                                                                                                              & 0                                                                   \\
\newline \\
SCENT      & Offests in log transformation                                                                                                                                                               & 1                                                                                            & Offests in log transformation                                                                                                                                                     & 1                                                                   \\
\newline \\
MCE        & None                                                                                                                                                                                        & 0                                                                                            & None                                                                                                                                                                              & 0                                                                   \\
\newline \\
SPIDE      & \begin{tabular}[c]{@{}l@{}}Offests in log transformation;\\ k in finding a cell's k-nearest \\ neighbors\end{tabular}                                                                       & 2                                                                                            & \begin{tabular}[c]{@{}l@{}}k in finding a cell's k-nearest \\ neighbors\end{tabular}                                                                                              & 1                                                                   \\
\newline \\
mRNAsi     & None                                                                                                                                                                                        & 0                                                                                            & None                                                                                                                                                                              & 0                                                                   \\
\newline \\
CCAT       & None                                                                                                                                                                                        & 0                                                                                            & None                                                                                                                                                                              & 0                                                                   \\
\newline \\
FitDevo    & Number of PCs                                                                                                                                                                                        & 1                                                                                            & Number of PCs                                                                                                                                                                               & 1   \\
\bottomrule
\end{tabular}
\end{adjustbox}
\label{tab:justify_tunable}
\end{table}
\clearpage

\section*{Supplementary Material}
\label{supp}
\beginsupplement

\begin{table}[h]
\begin{center}
\caption{\re{Inconsistencies} between publication and code implementation}\label{tab:discrepancy}

\begin{tabular}{lll}
\toprule
Method & Publication & Code Implementation\\
\midrule
\multirow{3}{*}{CytoTRACE$^\dagger$} & GCS $=$ Geometric Mean ($\tilde{Z}$) & GCS $=$ Arithmetic Mean ($\tilde{Z}$) \\
                            \newline \\
                           & No cell filtering is mentioned & Use $1{,}000$ genes to filter cells \\ \newline \\ \hdashline
\newline\\
NCG &$Z_{ij} = \log_2(S_{ij} + 1.1)$ for $S_{ij} < 1$ &$Z_{ij} = \log_2 1.1$ for $S_{ij} < 1$\\
\newline\\
stemID &  $E_j = \sum_i p_{ij} \log p_{ij}$ &  $E_j = -\sum_i p_{ij} \log p_{ij}$\\
\newline \\
dpath$^\ast$ & $\pi_{ij} = 0.1$ when $X_{ij} = 0$ & $\pi_{ij} = 0$ when $X_{ij} = 0$ \\
\newline \\
\hdashline
\newline \\
\multirow{6}{*}{SPIDE} 
&  \begin{tabular}[l]{@{}l@{}}cell-cell correlation is based on quantile \\normalized and log-transformed data\end{tabular}&  \begin{tabular}[l]{@{}l@{}}cell-cell correlation is based on\\ log-transformed data \end{tabular}\\
\newline \\
&$R_{ii^\prime j}$ is calculated using $Z$&\begin{tabular}[l]{@{}l@{}} $R_{ii^\prime j}$ is calculated using $X^\ast $\\ where $X^\ast_{ij} = \log_2(X_{ij} + 1.1)$ \end{tabular}\\
\newline \\
& $P_{ii^\prime j} = \frac{\lvert R_{ii^\prime j}\lvert Z_{i^\prime j}}{\sum_{i'' \in N(i)} \lvert R_{ii''j} \lvert^2 Z_{i''j}}$$^\ddagger$ & 
$P_{ii^\prime j} = \frac{\lvert R_{ii^\prime j}\lvert Z_{i^\prime j}}{\sum_{i'' \in N(i)} \lvert R_{ii''j} \lvert Z_{i''j}}$\\
\newline \\
                           & $\Pi_{ij} = \frac{Z_{ij}  (A \cdot Z_{\cdot j})_i}{Z_{\cdot j}^\intercal  \cdot  (A  \cdot  Z_{\cdot j})}$$^\S$ with undefined $A$& $\Pi_{ij} = \frac{Z_{ij}  (A_j \cdot Z_{\cdot j})_i}{Z_{\cdot j}^\intercal  \cdot  (A_j  \cdot  Z_{\cdot j})}$\\ 

\bottomrule
\end{tabular}
\end{center}
\footnotesize{$^\dagger$: Here $\tilde{Z}$ is the $\text{log}_2$-normalized expression matrix which includes the top $200$ genes whose expression levels have the highest correlations with the gene counts.}
\newline
\footnotesize{$\ast$: Here $\pi_{ij}$ denotes the weight for gene $i$ in cell $j$ used in the first round of weighted NMF. (Both rounds of weighted NMF use the notation $\pi_{ij}$, and the \re{inconsistency} only exists in the first round.)}
\newline
\footnotesize{$\ddagger$: This equation refers to equation $2$ in SPIDE's publication. We rewrote this equation using our unified notations.}
\newline
\footnotesize{$\S$: This equation refers to $\pi_{i} = \frac{x_i  (A x)_i}{(x^\intercal Ax)}$ in SPIDE's publication. We rewrote this equation using our unified notations.}
\end{table}
\newpage

\subsection*{Explanation of the inconsistency about the geometric mean and arithmetic mean in CytoTRACE}
We understand the fact that the arithmetic mean of logarithmic expression levels is equal to the logarithm of the geometric mean of (unlogarithmic) expression levels, provided the logarithm is taken without a pseudocount. However, this equality does not hold in CytoTRACE's algorithm, where a pseudocount is used in the log trasformation. Based on the following two quotes from CytoTRACE’s publication, it is reasonable to interpret that the geometric mean was calculated using logarithmic expression levels, with no mention of taking the logarithm after calculating the geometric mean.

    Quote 1: ``\quotes{The resulting gene expression matrix $X^\ast$ was log2-normalized with a pseudo-count of 1. We found that by using $X^\ast$ to derive GCS, as opposed to the original matrix $X$, we could obtain a larger gain in performance over gene counts alone (P = 0.006 vs. 0.04, respectively; paired two-sided t-test; fig. S8C). This finding prompted us to apply this transformation as a pre-processing step for GCS and CytoTRACE throughout this work.}''
    
    Quote 2: ``\quotes{Indeed, a gene counts signature (GCS), defined as the geometric mean of genes that are most correlated with gene counts, outperformed all evaluated features in the training cohort}.''
    
    Based on Quote 2, it is reasonable to interpret that the GCS for each cell is the geometric mean of the selected genes, whose expression levels can be interpreted as logarithmic based on Quote 1. However, this interpretation is inconsistent with the code, which defines GCS as the arithmetic mean of the logarithmic expression levels with a pseudocount. Moreover, the correspondence between the geometric mean and arithmetic mean no longer holds when a pseudocount is introduced. Therefore, our reported inconsistency is valid.

\clearpage
\begin{figure}
    \centering
    \includegraphics[width = \textwidth]{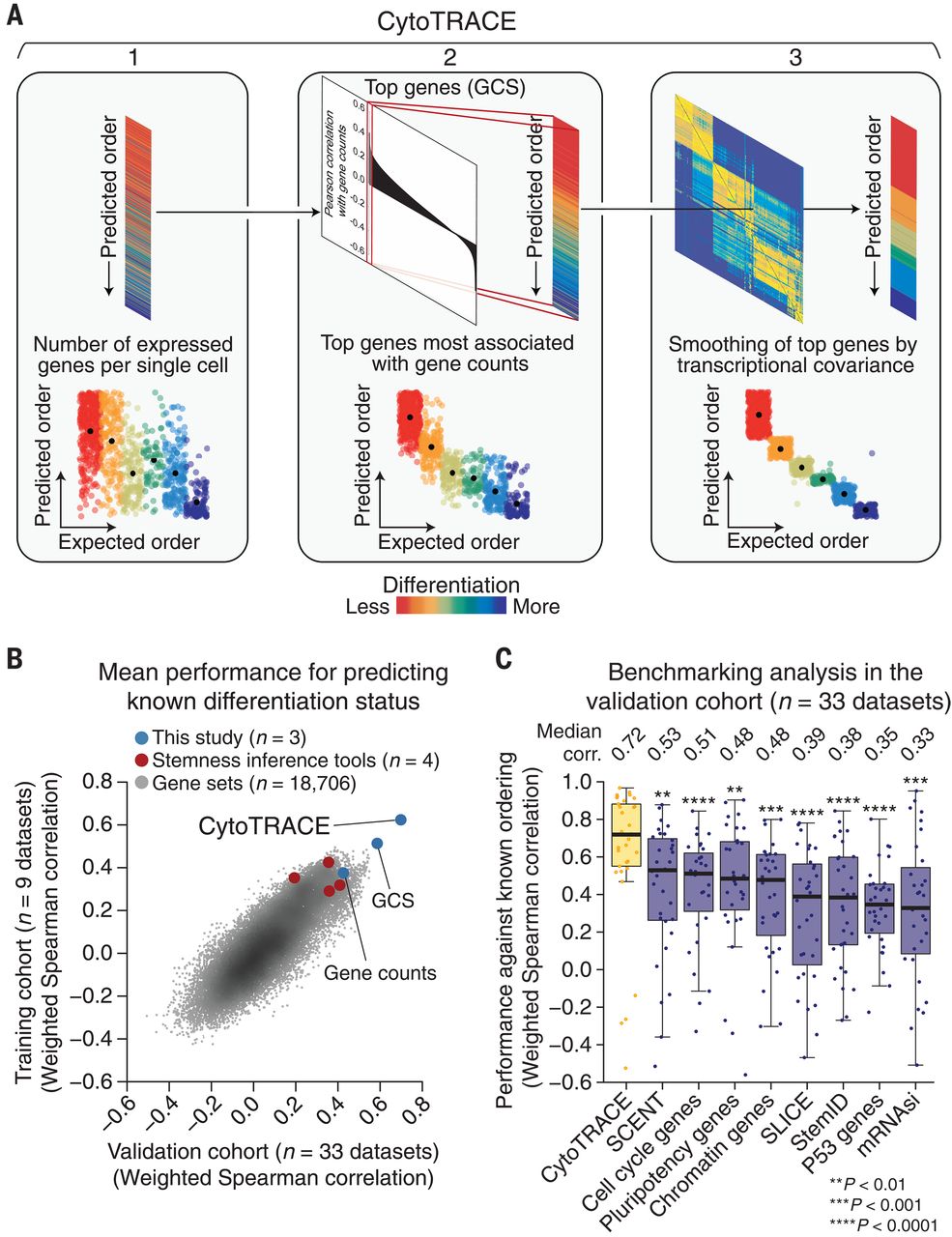}
    \caption{\re{This is Figure 2 from CytoTRACE's publication \cite{gulati2020cytotrace}, and permission will be purchased before the publication of this manuscript. Figure 2C shows that the performance of CytoTRACE and other computational methods can vary significantly across different datasets. In some cases, the Spearman correlation between the predicted cell potency and the ground truth can even be negative.}}
    \label{fig:cytotrace_2c}
\end{figure}

\begin{figure}
    \centering
    \includegraphics[width = \textwidth]{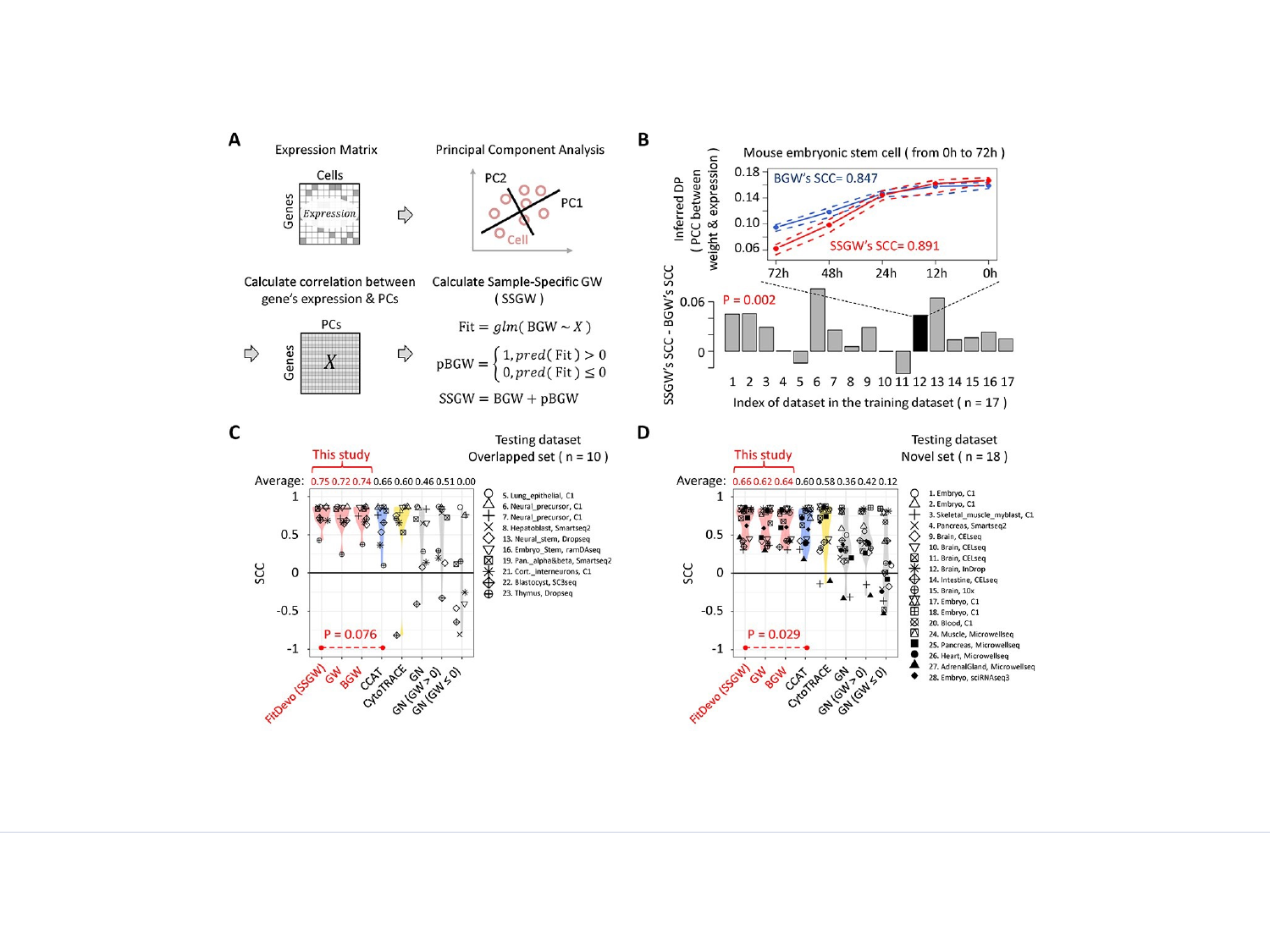}
    \caption{\re{This is Figure 3 from FitDevo's publication \cite{zhang2022fitdevo}, and this recreation is permitted under the open access terms. Figures 3C and 3D show that the performance of FitDevo, CCAT, and CytoTRACE varies to different degrees.}}
    \label{fig:fitdevo_3}
\end{figure}

\end{document}